\documentclass[paper,notoc]{JHEP3}

\usepackage{helvet}
\usepackage{amsmath}
\usepackage{amssymb}
\usepackage{setspace}
\usepackage[dvips]{graphicx}
\usepackage{appendix}
\usepackage{epsfig}

\def\be{\begin{equation}}
\def\ee{\end{equation}}

\author{L.A. Harland-Lang$^1$, V.A. Khoze$^{2,3}$, M.G. Ryskin$^{2,4}$, W.J. Stirling$^{1,2}$\\ 
  $^1$Cavendish Laboratory, University of Cambridge,
  J.J.\ Thomson Avenue, Cambridge, CB3 0HE, UK\\
  $^2$ Department of Physics and Institute for Particle Physics Phenomenology, University of Durham, DH1 3LE, UK\\
$^3$ School of Physics \& Astronomy, University of Manchester,
Manchester M13 9PL, UK
$^4$ Petersburg Nuclear Physics Institute, Gatchina, St. Petersburg, 188300, Russia}

\preprint{IPPP/09/70\\ DCPT/09/140 \\ Cavendish-HEP-09/17}

\title{Central Exclusive $\chi_c$ Meson Production at the Tevatron Revisited}

\abstract{Motivated by the recent experimental observation of exclusive $\chi_{c}$ events at the Tevatron, we revisit earlier studies of central exclusive scalar $\chi_{c0}$ meson production, before generalising the existing formalism to include $\chi_{c1}$ and $\chi_{c2}$ mesons. Although $\chi_{c0}$ production was previously assumed to be dominant, we find that the $\chi_{c1}$ and $\chi_{c2}$  rates for the experimentally considered $\chi_c \to J/\psi \gamma \to \mu^+ \mu^-\gamma$ decay process are in fact comparable to the $\chi_{c0}$ rate. We have developed a new Monte Carlo event generator, SuperCHIC, which models the central exclusive production of the three $\chi_c$ states via this decay chain, and have explored possible ways of distinguishing them, given that their mass differences are not resolvable within the current experimental set-up. Although we find that the severity of current experimental cuts appears to preclude this, the acceptance does not change crucially between the three states and so our conclusions regarding the overall rates remain unchanged. This therefore raises the interesting possibility that exclusive $\chi_{c1}$ and $\chi_{c2}$ production has already been observed at the Tevatron.}
\begin{document}

\section{Introduction}\label{intro}

The measurement of central exclusive production (CEP) processes
in high-energy proton -- (anti)proton collisions represents a 
very promising way to study the properties of new 
particles, from exotic hadrons to the Higgs boson, see for example
Refs.~\cite{DR}~-~\cite{Klempt}.

The CEP of an object $A$ may be written in the form
$$pp({\bar p}) \to p+A+p({\bar p}),$$ where $+$ signs are used to denote 
the presence of large rapidity gaps.
An attractive advantage of these reactions is that they provide an especially clean
environment in which to measure
the nature and quantum numbers (in particular, the spin and parity) of new states, see for example Refs.~\cite{Kaidalov03,CK,HKRSTW}.
A topical example is the CEP of
the Higgs boson~\cite{Khoze00}~-~\cite{epip}. This provides a novel and promising way to study 
in detail the 
Higgs sector at the LHC and gives a strong motivation
for the addition of near-beam
proton detectors to enhance the discovery and physics potential 
of the ATLAS and CMS detectors at the LHC \cite{FP420}~-~\cite{royon}.

\begin{figure}
\begin{center}
\includegraphics[height=13cm]{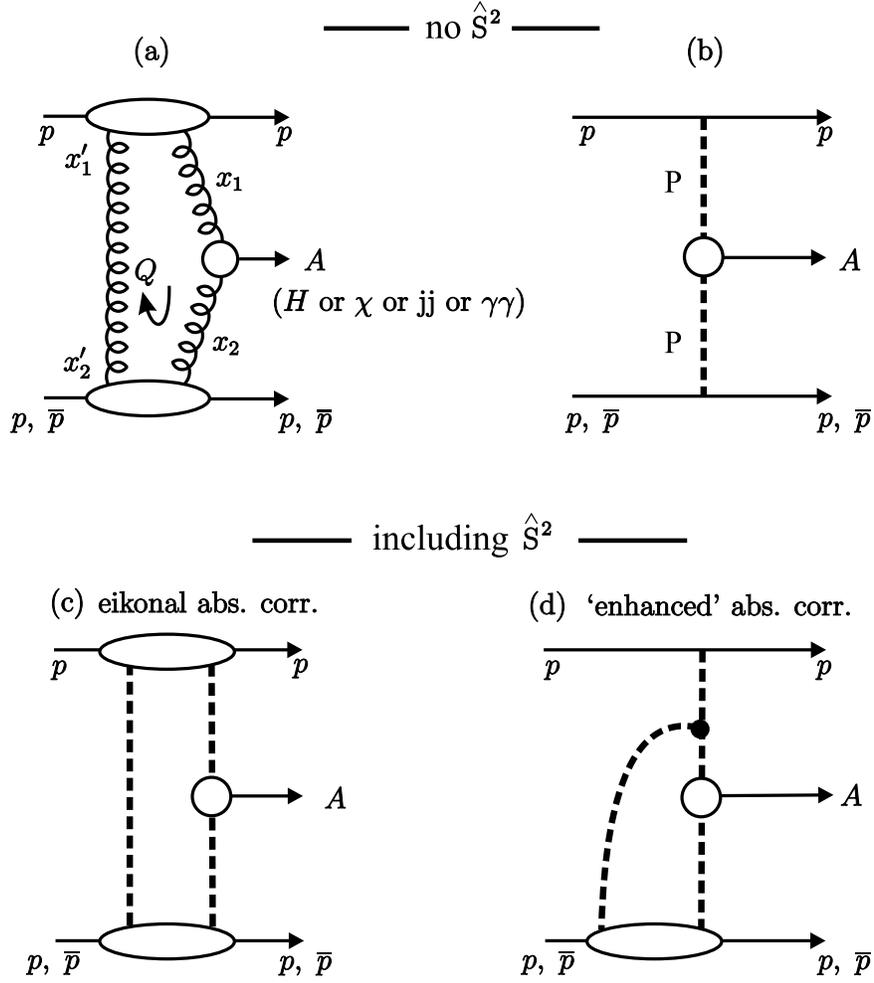}
\caption{Schematic diagrams for CEP of a system $A$
within the approach of Refs.~\cite{Khoze00,KMRprosp} and \cite{soft}~-~\cite{nns2}. 
The integration over the loop momentum $Q_\perp$ in diagram (a)
results
in a $J_z=0$  selection
rule~\cite{Khoze00a}, where $J_z$ is the projection of the total angular momentum
along the proton beam axis.
 It is also necessary
to compute the probability, ${\hat S}^2$, that the rapidity gaps survive soft 
(\cite{soft,nns1}) and semi-hard (\cite{JHEP}~-~\cite{bbkm}) rescattering;
these two possible types of unitarity (or absorptive) corrections
 are exemplified in diagrams (c) and (d) respectively, where the dashed lines
represent Pomeron exchanges (as in version (b) of diagram (a)).}  
\label{fig:sketch}
\end{center}
\end{figure}

Recently, exclusive diffractive processes $p\bar{p} \to p+A+\bar{p}$ have  been successfully 
observed by CDF Collaboration
at the Tevatron, where $A=\gamma\gamma$ \cite{CDFgg},
 dijet \cite{CDFjj} or $\chi_c$ \cite{Aaltonen09}.\footnote{For a recent review see \cite{Albrowrev}. More CDF exclusive data on $\gamma\gamma$ and $\chi_c$ production may be available in the near future~\cite{Albrow}.}
   As the sketch in 
Fig.~\ref{fig:sketch}(a)
 indicates, these processes
 are driven by the same mechanism as
exclusive Higgs (or other new object) production at the LHC,  
but have much larger cross sections. They can therefore serve as ``standard candles'', 
see \cite{KMRprosp,KMRSgg}, which allow us to check the theoretical
predictions for the CEP of new physics signals by measurements made
at the Tevatron. Moreover, the observed rates of all three CEP processes
measured by the CDF collaboration are in broad agreement with theoretical expectations
\cite{Khoze00a,Khoze04,KMRprosp,KMRSgg}, which 
lends credence to the predictions for exclusive
Higgs production at the LHC. 

Among the CEP processes measured at the Tevatron,
the double-diffractive production of C-even, heavy
quarkonia ($\chi_c$) states plays a special role \cite{Khoze04}
(see also \cite{Pump}~--~\cite{teryaev}). 
First, as is well known,
heavy quarkonium production provides  a valuable tool to test
the ideas and methods of the QCD 
 physics of bound states, such as  effective field
theories, lattice QCD, NRQCD, etc.
(see, for example, Refs.~\cite{Bodwin,Brambilla} for theoretical reviews).
Second, $\chi_c$ production exhibits characteristic features, based on Regge theory, that depend on the particle spin and parity $J^P$. We discuss these in Section~2 below. 

A potential problem with $\chi_c$ production as a ``standard candle" for Higgs production is that 
it is far from clear that the purely perturbative approach of  Refs.~\cite{Khoze00,KMRprosp}
 (as exemplified by Fig.~\ref{fig:sketch}(a)) is valid.  In particular,    the estimates
in  Ref.~\cite{Khoze04} assume  a perturbative contribution coming 
from integrating round the gluon loop in Fig.~\ref{fig:sketch}(a)
for $Q_\perp >0.85$~GeV. Due to such low scales, strictly speaking one
cannot guarantee that the accuracy in the perturbative predictions made in this way for 
$\chi_c$ CEP is better than a factor of 4-5 up or down. 
 In addition, there may be important non-perturbative contributions, traditionally modelled by the Pomeron-Pomeron process shown in Fig.~\ref{fig:sketch}(b), but here again there is  significant model dependence and so any predictions for this component also come with large uncertainties. We choose to take a pragmatic approach, in which we base our analysis on the perturbative contribution only, but at the same time we consider which features of the perturbative contribution (for example, the relative contributions of the various $J^P$ states, distributions of final state particles, etc.) are likely to be shared by the non-perturbative contribution.
Independent of the exact details of the production mechanism, the CEP of $\chi$ states provides
a valuable check on the important ingredients of the physics of Pomeron-Pomeron fusion.

Another important issue related to the CEP of $\chi_c$ states at the
Tevatron is that this process allows us to test the role of
the so-called enhanced absorptive corrections (see  Fig.~\ref{fig:sketch}(d)),
 which break soft-hard factorization, see for example \cite{nns2}~-~\cite{bbkm}.
As shown in \cite{epip,nns2},
there is a hierarchy in the value of the rapidity 
 gap survival factor $S$ due to enhanced absorption, $S_{\rm enh}$,
\begin{equation}\nonumber
S^{\rm LHC}_{\rm enh}(M_H>100 ~{\rm GeV})~~>~~S^{\rm Tevatron}_{\rm enh}(\gamma\gamma;E_T^\gamma>5 ~{\rm GeV})\\ \label{s:ineq}>~~S^{\rm Tevatron}_{\rm enh}(\chi_c),
\end{equation}
which reflects the size of the various rapidity gaps ($s/M^2$) of the different exclusive processes. 
The very fact that $\gamma\gamma$ and $\chi_c$ events have been observed at the Tevatron
in reasonable agreement with theoretical 
expectations\footnote{In the $\chi_c$ case the agreement becomes
especially striking after taking into account
the revised value of the total $\chi_{c0}$ width  which
has been reduced by a factor 1.4 \cite{PDG} as compared to the value
in the Review of Particle Properties (2002) used in  \cite{Khoze04}.}
 confirms
 that there is no danger that enhanced absorption will strongly 
reduce the exclusive SM Higgs signal at the LHC.

When interpreting the results of \cite{Aaltonen09}
in terms of the production of a particular  $\chi_c$ state
there is one important point to bear in mind.
While the $P$ and $C$ parities are unambiguously defined
by the fusion mechanism (see, for example, \cite {KMRprosp})
the spin assignment requires special care.
 As discussed in   Refs.~\cite{Khoze00a,Khoze04},
central exclusive $\chi_c$  production should be dominated by 
the $\chi_{c0}(0^{++})$ state. This is because $\chi_{c1}(1^{++})$ 
and $\chi_{c2}(2^{++})$
production is
{\it strongly suppressed}: the former
 due to the Landau-Yang theorem \cite{LY} for on-mass-shell gluons
and  the latter because in the non-relativistic
approximation
the $\chi_{c2}(2^{++})$ meson cannot be produced in
the $J_z=0$ state, which
dominates CEP for
forward outgoing protons \cite{Khoze00a}.

Recall,
however, that the experimental observation \cite{Aaltonen09} of exclusive $\chi_c$ production
is based on  the   
decay chain $\chi_c \to J/\psi \gamma \to \mu^+\mu^-\gamma$.
The observed
 $65 \pm 10$ signal events have a 
limited $M(J/\psi\gamma)$ resolution and are collected
in a restricted area of final state kinematics 
(due to cuts and event
selection criteria).
In order to determine the $\chi_c$ yield the dominance of $\chi_{c0}$ production
is {\it assumed} , and the CHIC Monte Carlo\footnote{CHIC is a publicly available
 Monte Carlo implementation of the $\chi_{c0}$ analysis of Ref.~\cite{Khoze04}.},
 based on the $\chi_{c0} \to J/\psi+\gamma$
decay, is used for conversion of the observed events into the cross section.
However at the present time we cannot rule out the possibility that, under the conditions
of the CDF experiment, higher spin $\chi_c$ states ($1^{++}$, $2^{++}$)  contribute to the 
observed $J/\psi+\gamma$ signal.\footnote{This applies not only to $\chi_c(1P)$ states, but
also to possible higher excitations $\chi_c(nP)$.}
As is correctly pointed out in   Ref.~\cite{teryaev},
the strong suppression of $1^{++}$ central production can 
be compensated by its 
much higher branching fraction to the $J/\psi+\gamma$ final state.
We show below (see also \cite{epip}) that this can also be true for
$\chi_{c2} (2^{++})$ CEP.
 $\,$ Explicitly, the $\chi_{c0},~\chi_{c1},~\chi_{c2}$ branching fractions to $J/\psi\gamma$ are
0.011, 0.34 and 0.19 respectively \cite{PDG}.
There is another factor leading to a further rebalance  between
the relative contributions of different $\chi_c$ spin states.
As discussed in \cite{Kaidalov03,epip},
the eikonal survival factor, $S_{\rm eik}$, is 
larger for $\chi_{c1}$ and $\chi_{c2}$ 
since, due to their spin structure, they are produced 
more peripherally.\footnote{ $S_{\rm enh}$ is largely independent of the
$\chi$ spin assignment.   
 Note also that the relative number of events where the forward protons
 dissociate is larger for $\chi_{c1}$ and $\chi_{c2}$ than for $\chi_{c0}$.}

The simultaneous presence of several $\chi_c$ states clearly requires
a more comprehensive analysis, including a 
new Monte Carlo programme, allowing for production and decay of the higher-spin states.
This is the main topic of the current paper, i.e.
we extend the analysis of  Refs.~\cite{Khoze00a,Khoze04}
to include the detailed study of $\chi_{c1}$ and $\chi_{c2}$
exclusive production. Special attention is paid
to the role of absorptive corrections, which significantly
affect the predicted rates, see \cite{Kaidalov03,Khoze04,epip}.

The paper is organized as follows. In Section~2 we review the general expectations for the $J^P$ 
properties of $\chi_c$ production based on Regge theory. In Section~3 we discuss in detail the perturbative approach to the calculation of $\chi_{c}$ CEP, paying particular attention to the differences between $\chi_{c0}$, $\chi_{c1}$ and $\chi_{c2}$ production and to the uncertainties in the predictions. We have implemented our calculations in a new Monte Carlo event generator -- SuperCHIC -- which is described in Section~4. In Section~5 we present numerical results for $\chi_{c0}$, $\chi_{c1}$ and $\chi_{c2}$ CEP at the Tevatron. We discuss the impact of the survival factors on the CEP cross sections, and comment on the size of possible non-perturbative contributions. We investigate to what extent kinematical distributions of the final-state particles can be used to distinguish the three $\chi_c$  spin states. In Section~6 we summarise our conclusions and  comment on possible future developments in the study of the CEP of quarkonia states at the Tevatron and LHC. Some additional calculational details are presented in two Appendices.

\section{General expectations from Regge theory}\label{Reggesec}

As discussed in \cite{Kaidalov03}, the central 
diffractive production
of meson states (see Fig.~\ref{fig:sketch}(b))
has characteristic features that depend on the particle spin and parity
$J^P$, which follow from the general principles of Regge theory.
Let us first recall 
particular examples of  {\it bare} Pomeron-Pomeron vertices 
 for the spin-parity $J^{P}$ of particle $h$  
 in the case of low transverse
momenta ${\bf p}_{1,2_\perp}$  of the outgoing protons.

\begin{itemize}
\item[(a)] $J^P(h) = 0^+$

For a scalar particle $h$, the vertex coupling is simply\footnote{We use boldface type to denote spatial three-vectors.}
\be g_{PP}^S = f_{0^+}({\bf p}_{1_\perp}^2,{\bf  p}_{2_\perp}^2, {\bf p}_{1_\perp}\!\cdot\!{\bf p}_{2_\perp}),
\label{eq:nightingale} \ee
where $f_{0^+}$ is a function of the displayed scalar variables.
 When ${\bf p}_{1_\perp}^2$ or ${\bf p}_{2_\perp}^2\to 0$, this
function, in general, tends to some constant $f_s$. 
Further information on the structure of
this function requires extra dynamical input. In particular,
within the perturbative framework (for ${\bf Q}_ {\perp}^2\gg {\bf p}_{1,2_\perp}^2$),
$f_{0^+}$ is almost independent of ${\bf p}_{1,2_\perp}$,
and hence the bare cross section, $\sigma_{0^+}$,
is essentially independent of the azimuthal
angle $\phi$  between the outgoing protons
\be
\frac{d\sigma_{0^+}}{dt_1dt_2d\phi} ~~\propto~~ {\rm constant}(\phi),
\label{eq:scal} \ee
where $t_{1,2}\simeq -{\bf p}_{1,2_\perp}^2$. We note, however, that for $\chi_c$ CEP the low $Q_\perp$ scale we are considering means that the inequality  ${\bf Q}_ {\perp}^2\gg {\bf p}_{1,2_\perp}^2$ is not
completely valid, and we therefore expect some deviation from the constant behaviour of (\ref{eq:scal}).

\item[(b)] $J^P(h) = 0^-$

For the central production of a pseudoscalar particle, the bare vertex factor takes the form
\be g_{PP}^P = f_{0^-}({\bf p}_{1_\perp}^2, {\bf p}_{2_\perp}^2, {\bf p}_{1_\perp}\!\cdot\! {\bf p}_{2_\perp})\:
({\bf p}_{1_\perp} \times {\bf p}_{2_\perp})\cdot {\bf n}, \label{eq:ocelot} \ee
where ${\bf n}$ is the unit vector in the direction of the colliding hadrons (in the c.m.s.).
Due to the identity of the Pomerons,
 the function $f_{0^-}$ should be symmetric under the interchange
$1\leftrightarrow 2$. It follows from  (\ref{eq:ocelot}) that in the pseudoscalar case
the bare cross section, $\sigma_{0^-}$, should behave for small $|t_{1,2}|$ as
\be \frac{d\sigma_{0^-}}{dt_1dt_2d\phi}~~\propto~~|t_1||t_2|\,{\rm sin}^2\phi .
\label{eq:pseud} \ee

An immediate consequence of  (\ref{eq:pseud}) is that pseudoscalar
production is forbidden when 
the protons scatter at zero angle.

\item[(c)] $J^P(h) = 1^+$

For the production of an axial vector state
the bare Pomeron-Pomeron fusion 
vertex factor can be written as
\be g_{PP}^A \!\sim\! a_{\lambda = 0}\frac{(t_1-t_2)([{\bf p}_{1_\perp} \times {\bf p}_{2_\perp}]\cdot {\bf e})}{M^2} +
a_{\lambda = 1} \frac{[{\bf K}\times {\bf n}]\cdot {\bf e}}{M} \label{eq:axial} \ee
with 
\be {\bf K}\equiv{\bf p}_1 -{\bf p}_2.
\ee
Here $M$ and ${\bf e}$ are the mass and polarization vector 
of the centrally produced $1^{++}$ state,
and
the vertex functions $a_{\lambda = 0,1}$ correspond to
axial meson production with helicities  $\lambda = 0,1$
in the target rest frame (where the longitudinal component of the axial meson momentum is much larger than its transverse component). Analogously to $f_{0^+}$ and $f_{0^-}$ in the previous cases, the functions $a_{\lambda=0,1}$ may depend on
 ${\bf p}_{1_\perp}^2,\; {\bf p}_{2_\perp}^2$ and $({\bf p}_{1_\perp}\cdot{\bf p}_{2_\perp})$, and are symmetric under the $1\leftrightarrow 2$ interchange.

It follows from (\ref{eq:axial}) that the bare amplitude 
tends to zero at low $K_{\perp}$, in particular when both protons scatter at zero angle. 
Another important consequence of (\ref{eq:axial}) is that at low $|t_{1,2}| $ the axial meson
should be produced dominantly in the helicity-one state.
As already mentioned in \cite{Kaidalov03}, the general structure of the
axial vertex $g_{PP}^A$, given by Eq.~(\ref{eq:axial}), coincides 
with that found using a non-conserved vector current model~\cite{CK}, which gives a
good description of the experimental data on $f_1(1285)$ and $ f_1(1420)$ CEP by
the WA102 Collaboration~\cite{a2}; for a review see Ref.~\cite{Klempt}.

\item[(d)] $J^P(h) = 2^+$

For a tensor particle $h$, the bare vertex function $g_{PP}^T$
is not constrained by Regge theory alone.
However, as we have already mentioned, within the perturbative approach of
Refs.~\cite{Khoze04,KMRprosp} the forward 
 CEP of non-relativistic heavy $2^{++}$  quarkonium
should be strongly reduced because of the
 suppression of the $2^{++}\to 2g$ transition
for the  $J_z=0$ on-mass-shell two-gluon state.\footnote{We reconfirm
the conclusion of \cite{Khoze00a} that the relativistic corrections
to the $\chi_{c,b}(2^{++})\to 2g$ transition are numerically small.}

\end{itemize}

\begin{figure}[h]
\begin{center}
\includegraphics[height=10cm]{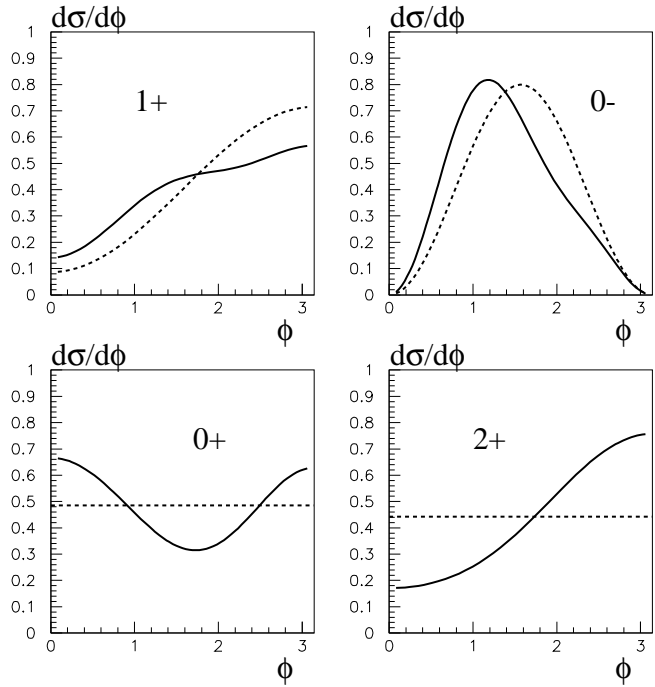}
\caption{Impact of the absorptive corrections on the distribution (in arbitrary units) 
of the difference in azimuthal angle of the outgoing protons for the CEP of various $J^P$ $\chi_c$ states at the LHC, using the two channel eikonal model of Ref.~\cite{soft}. The solid (dashed) lines are the distributions including (excluding) the survival factor. For completeness we also show the result for pseudoscalar $\eta_c$ production.}\label{schik}
\end{center}
\end{figure}

\begin{figure}[h]
\begin{center}
\includegraphics[scale=0.3]{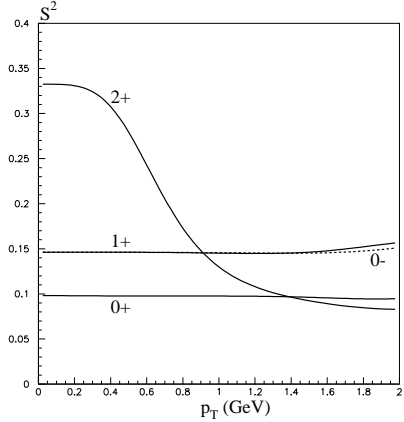}
\caption{Dependence of the survival factor $S^2_{\rm eik}$ on the transverse momentum of centrally produced $\chi_c$ mesons at the Tevatron, using the two channel eikonal model of Ref.~\cite{soft}. Also shown (dashed line) is the survival factor corresponding to $\eta_c$ production.}\label{schipt}
\end{center}
\end{figure}

Finally, we recall that,
as discussed in \cite{Kaidalov03}, the absorptive corrections
arising from the multi-Pomeron exchanges modify
the distributions over $\phi$. This is because the 
absorption depends on the distribution in impact parameter  ${\bf b}$
space, which in turn leads to a characteristic dependence
of the survival factor (mainly $S^{2}_{\rm eik})$ on the azimuthal angle between the outgoing protons. This effect was discussed in detail in~\cite{KMRforw}.

To demonstrate how the absorptive corrections may affect the angular distributions between the outgoing protons we have used a simple two 
channel eikonal model~\cite{soft}. We show in Fig.~\ref{schik} the results for $\sqrt s=14$~TeV. 
An analogous dependence of $S^2_{\rm eik}$ on the transverse momentum of the centrally produced meson is shown for the Tevatron energy $\sqrt s=1.96$~TeV 
in Fig.~\ref{schipt}. 
 
\section{Central Exclusive $\chi_c$ production: perturbative framework}\label{theory}
\begin{figure}
\begin{center}
\resizebox{0.4\textwidth}{!}{\includegraphics[]{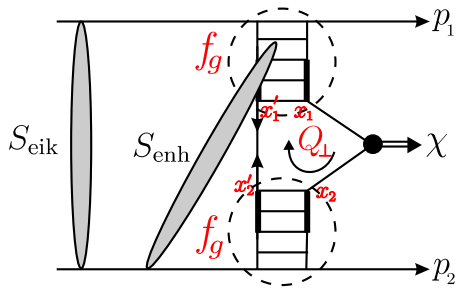}
}
\caption{The perturbative mechanism for the exclusive process $pp \to p+\chi+p$, with the eikonal and enhanced survival factors 
shown symbolically.}
\label{fig:pCp}
\end{center}
\end{figure}

To calculate the perturbative contribution to the central exclusive $\chi_c$ production process we use the formalism of Refs.~\cite{Khoze00a,Kaidalov03,Khoze00}. The amplitude is described by the diagram shown in Fig.~\ref{fig:sketch}(a), 
where the hard subprocess $gg \to \chi_c$ is initiated by gluon-gluon fusion and the second $t$-channel gluon is needed to screen the colour flow across the rapidity gap intervals. We can write the Born amplitude in the factorised form~\cite{Khoze04,KKMRext} (see Fig.~\ref{fig:pCp}):
\begin{equation}\label{bt}
T=\pi^2 \int \frac{d^2 {\bf Q}_\perp\, V_J}{{\bf Q}_\perp^2 ({\bf Q}_\perp-{\bf p}_{1_\perp})^2({\bf Q}_\perp+{\bf p}_{2_\perp})^2}\cdot f_g(x_1,x_1', Q_1^2,\mu^2;t_1)f_g(x_2,x_2',Q_2^2,\mu^2;t_2) \; ,
\end{equation}
where $V_J$ is the colour-averaged, normalised sub-amplitude for the $gg \to \chi_{cJ}$ process:
\begin{equation}\label{Vnorm}
V_J\equiv \frac{2}{s}\frac{1}{N_C^2-1}\sum_{a,b}\delta^{ab}p_1^\mu p_2^\nu V_{\mu\nu}^{ab} \; .
\end{equation}
Here $a$ and $b$ are colour indices and $N_C=3$. The amplitude $V_{\mu\nu}^{ab}$ represents the coupling of two gluons to the $\chi_c$ state being considered: the procedure for calculating this is outlined below.
The $f_g$'s in (\ref{bt}) are the skewed unintegrated gluon densities of the proton at the hard scale $\mu$, taken typically to be of the order of the produced massive state, i.e. $M_\chi/2$ in this case, and only one transverse momentum scale is taken into account by the prescription
\begin{align}\nonumber
Q_1 &= {\rm min} \{Q_\perp,|({\bf Q_\perp}-{\bf p}_{1_\perp})|\}\\ \label{minpres}
Q_2 &= {\rm min} \{Q_\perp,|({\bf Q_\perp}+{\bf p}_{2_\perp})|\} \; .
\end{align}
The longitudinal momentum fractions carried by the gluons satisfy
\begin{equation}\label{xcomp}
\bigg(x' \sim \frac{Q_\perp}{\sqrt{s}}\bigg)  \ll \bigg(x \sim \frac{M_\chi}{\sqrt{s}}\bigg) \; .
\end{equation} 
The $t$ dependence of the $f_g$'s is not well known, but in the limit that the protons scatter at small angles, we can assume a factorization of the form
\begin{equation}
f_g(x,x',Q^2_i,\mu^2;t)=f_g(x,x',Q^2_i,\mu^2)\,F_N(t) \; ,
\end{equation}
where the $t$-dependence is isolated in a proton form factor, which we take to have the phenomenological form $F_N(t)={\rm exp}(bt/2)$. In such a case a fit to soft hadronic data~\cite{soft} gives $b \simeq 4\, {\rm GeV}^{-2}$, which is also consistent with extracting $F_N(t)$ from the $t$-dependence in `elastic' $J/\psi$ photoproduction~\cite{Aid96}. We will therefore use $b=4\,{\rm GeV}^{-2}$ as our value for the slope parameter throughout.

In the kinematic region specified by (\ref{xcomp}), the skewed unintegrated densities are given in terms of the conventional (integrated) densities $g(x,Q_i^2)$. To single log accuracy, we have\footnote{In actual calculations, we use a more precise phenomenological form given by Eq.~(26) of~\cite{Martin01ms}.}
\begin{equation}\label{fgskew}
f_g(x,x',Q^2_i,\mu^2)=R_g\frac{\partial}{\partial \, {\rm log} \, Q^2_i} \big[x g(x,Q^2_i)\sqrt{T_g(Q^2_i,\mu^2)}\big] \; ,
\end{equation}
where $T_g$ is the usual Sudakov survival factor which ensures that the active gluon does not emit additional real partons in the course of the evolution up to the hard scale $\mu$, so that the rapidity gaps survive. $R_g$ is the ratio of the skewed $x' \ll x$ unintegrated gluon distribution to the conventional diagonal density $g(x,Q^2)$. For $x \ll 1$ it is completely determined~\cite{Shuvaev99}. The explicit form for $T_g$ is given by resumming the virtual contributions to the DGLAP equation. It is given by
\begin{equation}\label{ts}
T_g(Q_\perp^2,\mu^2)={\rm exp} \bigg(\!\!-\!\!\int_{Q_\perp^2}^{\mu^2} \frac{d {\bf k}_\perp^2}{{\bf k}_\perp^2}\frac{\alpha_s(k_\perp^2)}{2\pi} \int_{0}^{1-\Delta} \!\bigg[ z P_{gg}(z) + \sum_{q} P_{qg}(z) \bigg]dz \!\bigg) .
\end{equation}
Here, as in~\cite{Khoze04}, we go beyond the collinear approximation and in the $T$ factor we resum not just the single collinear logarithms, but the single soft $\ln(1-z)$ terms as well. To a good approximation, this can be achieved by taking the upper limit of the $z$ integration in (\ref{ts}) to be
\begin{equation}\label{delta}
\Delta=\frac{k_\perp}{k_\perp+0.62M_\chi} \; .
\end{equation}
Returning to the $gg \to \chi_{c}$ amplitude, we note that the original extension of the CEP formalism to $\chi_{c0}$ production in~\cite{Khoze04} was achieved in direct analogy to the Higgs case, that is  by assuming that the $\chi_{c0}$ coupled to the gluons as a pure scalar with any effects from its internal structure neglected. We now wish to go beyond this approximation and model the internal structure of the $\chi_c$ meson for all three $J$ states and in particular their coupling to two gluons. This is done by a simple extension of the calculation of~\cite{Kuhn79}, where the coupling of ${}^3 P_J$ quarkonium states to two off-mass-shell photons is considered: as the gluons are in a colour singlet state the only difference will be constant prefactors resulting from colour algebra. 
We will simply state the results for the three amplitudes, leaving the derivation to Appendix \ref{chis}:
\begin{align}\label{V0}
V_0&=\sqrt{\frac{1}{6}}\frac{c}{M_\chi}((q_{1_{\perp}}q_{2_{\perp}})(3M_\chi^2-q_{1_{\perp}}^2-q_{2_{\perp}}^2)-2q_{1_{\perp}}^2q_{2_{\perp}}^2)  \; ,
\\ \label{V1}
V_1&=-\frac{2ic}{s} p_{1,\nu}p_{2,\alpha}((q_{2_\perp})_\mu(q_{1_\perp})^2\!-\!(q_{1_\perp})_\mu(q_{2_\perp})^2)\epsilon^{\mu\nu\alpha\beta}\epsilon^{*\chi}_\beta  \; ,\\ \label{V2}
V_2&=\frac{\sqrt{2}cM_\chi}{s}(s(q_{1_\perp})_\mu(q_{2_\perp})_\alpha+2(q_{1_\perp}q_{2_\perp})p_{1\mu}p_{2\alpha})\epsilon_\chi^{*\mu\alpha}  \; ,
\end{align}
where $q_{1_{\perp}}\equiv Q_\perp - p_{1_\perp}$ and $q_{2_{\perp}}\equiv -Q_\perp - p_{2_\perp}$. The amplitudes are normalised as in (\ref{Vnorm}) and the $q_{i_\perp}$ are $4$-vectors with $q_{i_{\perp}}^2 \equiv -{\bf q}_{i_{\perp}}^2<0$ throughout.\footnote{Four-vector scalar products are denoted by $(pq)$.}
 Considering first the $\chi_{c0}$ vertex, in the ${\bf Q}_\perp^2 \ll M_\chi^2$ limit (which is true to an acceptable degree of accuracy) we expect the internal structure of the $\chi_{c0}$ to be unimportant, and therefore to recover the previous result of~\cite{Khoze04}. We find
\begin{equation}\label{v0norm}
V_0 \approx \frac{48\pi \alpha_S}{\sqrt{N_C} M_\chi^3}\frac{\phi_c'(0)}{\sqrt{\pi M_\chi}}(q_{1_{\perp}}q_{2_{\perp}}) \; .
\end{equation}
Making use of the standard NRQCD result (see for example Refs.~\cite{Barbieri:1975am,Novikov:1977dq,Close97}),
\begin{equation}
\Gamma(\chi_{c0} \to gg)=96\frac{\alpha_S^2}{M_\chi^4}|\phi_c'(0)|^2 \; ,
\end{equation}
we find\footnote{Analogously to Ref.~\cite{Khoze04}, we assume the same NLO correction for the $gg \to \chi$ vertex as for the $\chi \to gg$
width, which can be valid only within a certain approximation. Moreover, as has been known for some time in the $P-$wave case (see for example
Ref.~\cite{Barbieri:1980yp}), the NNLO and higher-order radiative corrections to the $\chi \to gg$ transition are expected to be numerically quite large,
and this would result in further uncertainties in the theoretical expectations. Recall that these corrections are not
universal and depend on the spin-parity assignment of the $P-$wave states. In particular, it is known that in the $\chi_0$ case the dominant part of
the NLO correction comes from the $(i\pi)^2$ term originating from the Sudakov-like double logarithm $\alpha_s\ln^2(q^2/M^2)$, when the
imaginary part of the logarithm ($=-i\pi$) is squared. This double logarithm, and correspondingly the $(i\pi)^2$ contributions, are absent
for the case of the $\chi_1$, where the amplitude for on-mass-shell ($q^2=0$) gluons vanishes due to the Landau-Yang theorem.}
\begin{equation}
|V_0|^2=\frac{8\pi \Gamma(\chi_{c0}\to gg)}{M_\chi^3}(q_{1_{\perp}}q_{2_{\perp}})^2  \; ,
\end{equation}
which has the same form and normalisation as the previous result, as it must do. We will take this large $M_\chi$ limit throughout. Turning now to the $\chi_{c1}$ vertex, we can immediately see that it vanishes for on-shell gluons, that is when $q_i^2=q_{i\perp}^2=0$, as dictated by the Landau-Yang theorem (see Section~2). Furthermore, in the forward limit we have $q_{1\perp}=-q_{2\perp}=Q_\perp$ and so
\begin{align}
V_0 &\to -\sqrt{\frac{3}{2}}c M_\chi Q_\perp^2\; ,\\ 
V_1 &\to \frac{4ic}{s} Q_\perp^2 p_{1,\nu}p_{2,\alpha}Q_{\perp \mu} \epsilon^{\mu\nu\alpha\beta}\epsilon^{*\chi}_\beta\; ,\\ 
V_2 &\to -\frac{\sqrt{2}cM}{s}(sQ_{\perp\mu}Q_{\perp\alpha}+2Q_{\perp}^2p_{1\mu}p_{2\alpha})\epsilon_\chi^{*\mu\alpha} \; .
\end{align}
We see that $V_1$ is odd in $Q_\perp$, and will therefore vanish upon the loop integration (\ref{bt}) over ${\mathbf Q_\perp}$. For $V_2$ we make use of the identity
\begin{equation}\label{Qang}
\int d^2Q_\perp Q_{\perp\mu}Q_{\perp\sigma}=\frac{\pi}{2}\int dQ^2_\perp Q_\perp^2 g_{\mu\sigma}^{_T} \; ,
\end{equation}
where $g_{\mu\sigma}^{_T}$, the transverse part of the metric, can be written in the covariant form
\begin{equation}
g_{\mu\sigma}^{_T}=g_{\mu\sigma}-\frac{2}{s}(p_{1\mu}p_{2\sigma}+p_{1\sigma}p_{2\mu}) \; .
\end{equation}
We then find $V_2 \propto \epsilon_{\phantom{\mu}\mu}^{\mu}$ which vanishes due to the tracelessness of the $\chi_2$ polarization tensor (\ref{eps2}). We see that, as expected, the $\chi_{c2}$ and $\chi_{c1}$ production amplitudes vanish in the forward limit, and we will therefore expect the corresponding rates to be suppressed relative to $\chi_{c0}$ production, via the integration over the proton form factor $e^{b t_i} \approx e^{-b {\mathbf p_{i\perp}^2}}$ which suppresses large ${\mathbf p_{i\perp}^2}$ values. In fact we can give a very rough estimate for the level of suppression we will expect. 
Squaring and summing over polarization states gives
\begin{equation}\label{comp}
|V_0|^2:|V_1|^2:|V_2|^2 \sim 1:\frac{\left\langle \mathbf{p}_{\perp}^2\right\rangle}{M_\chi^2}:\frac{\left\langle \mathbf{p}_{\perp}^2\right\rangle^2}{\left\langle \mathbf{Q}_\perp^2\right\rangle^2} \; .
\end{equation}
Note that this result, as well as the amplitudes of (\ref{V0} - \ref{V2}), is also applicable to the CEP of $\chi_b$ mesons. The factor of $\left\langle \mathbf{p}_{\perp}^2\right\rangle$ comes from integrating over the assumed exponential form of the proton vertex,
\begin{equation}
\left\langle \mathbf{p}_\perp^2 \right\rangle = \int d\mathbf{p}_\perp^2 e^{-b\mathbf{p}_\perp^2} =\frac{1}{b} =\frac{1}{4}\, {\rm GeV}^{2} \; .
\end{equation}
If for simplicity we assume $\left \langle \mathbf{Q}_\perp^2 \right\rangle\approx 1.5\, {\rm GeV}^2$ and \linebreak[4] $M_\chi^2\approx 10\, {\rm GeV}^2$, 
we obtain\footnote{In fact, accounting for the values of $p_{1\perp},
p_{2\perp}$ in the denominator of (3.1) we obtain a slightly larger value of
the effective slope $b^{\rm eff}=b+O(1/Q^2_\perp)>4$ GeV$^{-2}$.
Therefore we expect a slightly larger suppression of the higher spin,
$\chi_1$ and $\chi_2$, states than that given by (3.22). Note also that
   for a heavier meson ($\chi_b$) CEP the slope
$b^{\rm eff}$ will be smaller than that for the case of $\chi_c$ due to a
typically larger values of $Q_\perp$ in the integral (3.1).}
\begin{equation}\label{rough}
|V_0|^2:|V_1|^2:|V_2|^2 \sim 1:\frac{1}{40}:\frac{1}{36} \; .
\end{equation}
While it is clear that we will have a quite sizeable suppression of the $\chi_{c1}$ and $\chi_{c2}$ CEP cross sections, these values are of course only very rough estimates, and an explicit calculation is required to confirm them.

We stress again that the legitimacy of the extension of the purely perturbative QCD treatment for central exclusive Higgs production to the $\chi_{c}$ case is somewhat questionable. 
For Higgs production the hard scale $\mu$ is set by $M_H/2$, and so we expect that a reliable calculation within perturbative QCD can be performed. In particular, the Sudakov factor leads to an IR stable result, with only a small contribution to the cross section from the region of $Q_\perp$ below $\sim 1$ GeV (although this is not to say the calculation does not come with significant uncertainties). However, in the case of $\chi_{c}$ production, where the `hard' scale is $\sim 1$ GeV, we expect and find that a significant part of the cross section comes from the IR unstable low $Q_\perp$ region. 
It might seem then, that despite the attractions of considering central exclusive $\chi_c$ production, any attempt to calculate a reliable cross section within the perturbative QCD framework is unlikely to succeed. This is not the case: the philosophy we take is that, even if the purely perturbative calculation is not IR stable, we should expect a smooth matching between the perturbative regime and the `soft' regime to which we can apply a non-perturbative Regge model. 
This was done in~\cite{Khoze04} for the $\chi_{c0}$, with a simple model for the Pomeron invoked, and the non-perturbative and perturbative contributions were found to be of a similar size, which gives justification for the inclusion of a perturbative contribution to $\chi_c$ CEP. On the other hand, there is much uncertainty surrounding which non-perturbative models for the Pomeron are most appropriate in this context, and so results such as this can only be used as a guide. 

Returning to the perturbative calculation, several comments are in order. As mentioned above, in the case of the $\chi_c$ we expect a significant proportion of the cross section to come from the low $Q_\perp$ region where perturbation theory is not valid, and there will correspondingly be a large degree of uncertainty in its predicted value. Given our lack of detailed understanding of low $Q_\perp$ non-perturbative gluon dynamics, the best we can do is to introduce an infrared cut-off to the $Q_\perp$ loop integral such that we are only considering the regime where perturbation theory will be reliable. This can be loosely justified on the grounds that our current understanding of these non-perturbative dynamic predicts that the low $Q_\perp$ contribution appears to be suppressed~\cite{Alkofer03}. Nevertheless significant uncertainties remain, both in the specific choice of cut-off, for which we have rough physical guidelines but which inevitably amounts to a subjective decision, and in what contribution we would actually expect from the low $Q_\perp$ region, which a cut-off prescription simply ignores.

\begin{figure}\begin{center}
\includegraphics[scale=0.6]{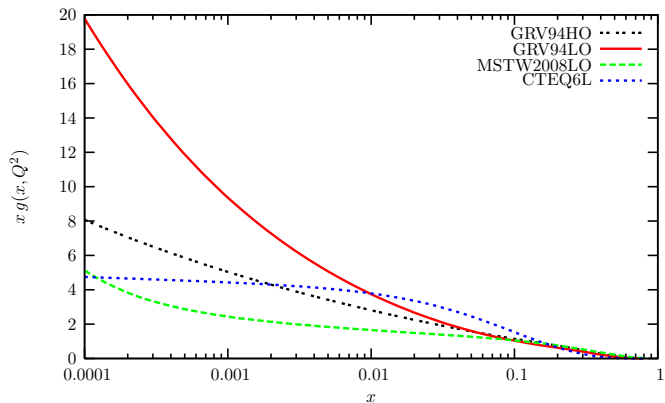}
\caption{LO and NLO PDFs at $Q^2=1.5\,$ GeV$^2$, plotted as function of $x$. Four representative PDF sets are displayed, and a large uncertainty at small $x$ is clear.} 
\label{PDFcomp}
\end{center}\end{figure}

A further uncertainty that is worth mentioning arises from the skewed PDFs, which we can express via (\ref{fgskew}) in terms of the conventional PDFs $g(x,Q^2)$, and therefore evaluate. Unfortunately, although this is in principal true, there is a large degree of uncertainty in the value of the conventional PDFs at the low $x$ and low $Q^2$ scales we are considering, as can be seen in Fig.~\ref{PDFcomp} where four representative PDF sets are plotted at $Q^2= 1.5\,$ GeV$^2$. In the case of exclusive $\chi_c$ production at the Tevatron ($\sqrt{s}=1.96$~TeV) we are sampling the $x \sim 2 \times 10^{-3}$ region, for which there is a large uncertainty. Recalling that the final cross section depends quartically on the skewed PDFs, this is clearly unacceptable. We can see, on the other hand, that as $x$ is increased the uncertainty rapidly decreases and we have a reasonable agreement between the sets in the region $0.05 \lesssim x \leq 1$. We can therefore perform the cross section calculation for a lower c.m.s energy such that the sampled $x$ value is in this range, although we must be careful that the corresponding $x$ value is not too high, as our initial formula for the skewed PDFs relies on a small $x$ approximation. In actual calculations a value of $\sqrt{s}=60\,$ GeV is chosen, which corresponds to $x \sim 0.05$. To make contact with the experimental c.m.s. energies of the Tevatron we then assume that the total cross section exhibits the Regge behaviour
\begin{equation}
d\sigma \propto s^{\alpha_P(t_1)+\alpha_P(t_2)-2} \; .
\end{equation}
This gives us a simple way to avoid the large uncertainties of the PDFs in the low $x$ region, although our final result will depend on the validity of this Regge assumption and on the specific value of the Pomeron intercept $\alpha_P(0)$ that we use.\footnote{Our procedure is equivalent to assuming a Regge-based extrapolation of the PDFs to small $x$.}

We end this section with a brief review of the uncertainties that are present in our calculation. First, we have the uncertainty in our choice of hard scale $\mu$ and the prescription (\ref{minpres}) for the transverse momentum scale $Q_i^2$, with for example the choice $\mu=M_H/4$ in the Higgs case giving a quoted 
enhancement of 30\% to the cross section~\cite{Kaidalov03}. Further to this we have the even more considerable uncertainty coming from the dependence of the cross section on the conventional PDFs to the fourth power in the low $x$ and $Q^2$ region, where they are not well determined. 
We also have the dependence of the final result on the non-perturbative, non-universal survival factor, 
which gives perhaps the largest uncertainty in the overall production rate. Differences in the higher-order QCD radiative corrections to the
$gg \to \chi$ vertex could also cause additional uncertainties. 
Finally we have the uncertainty in the low $Q_\perp$ contribution to the perturbative amplitude, which we find to be quite large in the $\chi_c$ case. For more details of these issues we refer the reader to~\cite{Kaidalov03}.

\section{SuperCHIC Monte Carlo Generator}\label{MC}
In~\cite{Khoze04}, central exclusive $\chi_{c0}$ meson production and its subsequent decay to $J/\psi \gamma \to \mu^+ \mu^- \gamma$ was modelled using the CHIC Monte Carlo event generator, but we now wish to consider the case where higher spin $\chi_c$ states are produced. The CEP of the three $\chi_{c(0,1,2)}$ states is modelled using a new, more general, Monte Carlo programme, SuperCHIC.\footnote{The extension to, for example, the CEP of pseudo-scalar $\eta_c$ or higher excitation $\chi_{c}(nP)$ states, as well as the respective $b$-quark states $\chi_{b}$, $\eta_b$ and $\chi_{b}(nP)$, is planned for future work.}  This follows essentially the same procedure for generating the relevant phase space as the previous CHIC MC -- optimised to reduce the event weight variation -- but with some important generalisations included that we will now outline.

The explicit evaluation of (\ref{bt}) is in all cases performed `offline' from the SuperCHIC  Monte Carlo event generator: to perform the loop integration for each event would lead to an unacceptably large run-time. The skewed PDFs are calculated as outlined in Section \ref{theory}, with the Sudakov factor read in from a grid to minimise run-time, and (\ref{bt}) then evaluated using standard Monte Carlo techniques.
We go beyond the approximation used in~\cite{Khoze04}, where the amplitude squared was calculated in the forward limit, with the $p_\perp$ dependence isolated in the proton form factors. Such an assumption is not relevant for the Tevatron, where the $p_\perp$ of the final state protons is not measured~\cite{Aaltonen09}, and so we must include non-forward effects. This will not only give a more accurate evaluation of the $\chi_{c0}$ cross section, but is also essential in the case of $\chi_{c1}$ and $\chi_{c2}$ production, where the corresponding amplitudes vanish in the forward limit. On the other hand, we still need to perform the loop integration separately from the Monte Carlo event generator. Considering first the $\chi_{c0}$ amplitude $A_0$, this can be achieved by noting that for small $p_\perp$ it must have the form
\begin{align}\nonumber
A_0 &\propto \int \frac{d^2 Q_\perp ({\bf q}_{1_{\perp}}\!\cdot\!{\bf q}_{2_{\perp}})}{Q_\perp^2 {\bf q}_1^2 {\bf q}_2^2}\,f_g(x_1,Q_1^2,\mu^2)f_g(x_2,Q_2^2,\mu^2)\\ \label{taylor0}
&\approx C_0+C_1({\mathbf p^2_{1_{\perp}}}+{\mathbf p^2_{2_{\perp}}})+C_{12}({\mathbf p_{1_{\perp}}}\cdot{\mathbf p_{2_{\perp}}})+\cdots \; ,
\end{align}
that is, there exists a Taylor expansion for $A_0$ formed from all possible scalar combinations of the $p_{i_\perp}$, the validity of which depends on the suppression in $p_\perp^2$ coming from the proton form factors. 
Squaring (\ref{taylor0}) and keeping only the leading terms in $p_{i_\perp}^2$, we can see that this expansion is equivalent to making the replacement (at lowest order in $p_{i_\perp}^2$)
\begin{equation}\label{exp}
e^{-b{\mathbf p^2_{i_{\perp}}}} \to e^{-(b-2\frac{C_1}{C_0}){\mathbf p^2_{i_{\perp}}}} \; .
\end{equation}
Thus to a first approximation we expect the inclusion of non-zero $p_\perp$ in the amplitude calculation to simply result in a change in the effective slope of the proton form factor. This then allows for an easy way to take into account the effect of non-forward protons in the amplitude, as we can simply model $|A_0|^2$ as a Gaussian, 
\begin{equation}\label{gfit0}
e^{-b({\mathbf p^2_{1_\perp}}+{\mathbf p^2_{2_\perp}})}\,|A_0|^2 \propto e^{-b^{\rm eff}_0({\mathbf p^2_{1_\perp}}+{\mathbf p^2_{2_\perp}})} \; ,
\end{equation}
where the slope $b^{\rm eff}_0$ and the overall normalisation are set by matching the values of $\langle p_{\chi_\perp}^2 \rangle$ and the integrated cross section, respectively, to those given by the exact expression for $|A_0|^2$. We note that the exact expression for $A_0$ (and therefore the $p_{\chi_\perp}$ distribution) will depend in general on the azimuthal angle between the outgoing protons, but by choosing to model this effect by simple Gaussians in $p_{i_\perp}^2$ the resultant azimuthal correlations between the protons will not be fully modelled in the Monte Carlo. Thus in the case of the $\chi_{c0}$ any deviation, for example, from the flat behaviour of (\ref{eq:scal}) is ignored. On the other hand it is clear that we are not currently interested in correctly modelling the $p_\perp$ distributions of the outgoing protons, which we recall are not measured at the Tevatron, but only those of the centrally produced final state particles, which it is important to know when discussing possible methods for distinguishing the three $\chi_c$ states, and the simple Gaussian approximation achieves this to an acceptable degree of accuracy.\footnote{In fact, in the case of the $\chi_{c2}$ it is necessary to include a $\mathbf{p_{1_\perp}}\cdot \mathbf{p_{2_\perp}}$ term in the fit. We also note that, while it is not done here, the complete inclusion of the correct azimuthal correlations, which may be relevant for
measurements at the LHC with tagged forward protons, remains a possible future extension of the Monte Carlo.}
However, we should be careful in our application of this approximation, as no strict $Q_\perp \gg p_\perp$ hierarchy exists for the $\chi_c$.\footnote{Note that such a procedure can be useful in the case
of CEP of the Higgs boson \cite{KKMRext,KMRprosp}.}

The higher $J$ states are more complicated due to their non-trivial Lorentz structure, but the basic argument remains the same. We can write the $\chi_{c1}$ and $\chi_{c2}$ amplitudes, omitting the $\chi_c$ polarization vectors etc. for simplicity, as
\begin{align}\label{a1fit}
A_1^\mu &\propto (p_{2_\perp}-p_{1_\perp})^\mu e^{-{b^{\rm eff}_1}({\mathbf p^2_{1_\perp}}+{\mathbf p^2_{2_\perp}})/2}
\; ,\\ \label{a2fit}
A_2^{\mu\nu} \!&\propto \!(s(p_{1\perp})^\mu(p_{2\perp})^\nu\!+\!2(p_{1\perp}p_{2\perp})p_1^\mu p_2^\nu)e^{-{b^{\rm eff}_2}({\mathbf p^2_{1_\perp}}+{\mathbf p^2_{2_\perp}})/2} \; ,
\end{align}
where we must now square the amplitudes and sum over the relevant $\chi_c$ polarization states before matching the values of the normalisation and slope as before.

SuperCHIC is a standard MC event generator that calculates the relevant weight for each generated event using these effective slopes, which can simply be read in at the beginning of the run. We give the option of generating events for purely $\chi_{c0}$, $\chi_{c1}$ and $\chi_{c2}$ production, as well as the (experimentally relevant) option of generating all three states at once. It is possible to generate the differential cross section ${\rm d}\sigma/{\rm d}y_\chi$ at a given rapidity $y_\chi$ value or the full cross section over a pre-specified $\chi_c$ rapidity range, with an approximate phenomenological fit invoked for the $y_\chi$ dependence of the cross section.

Having generated the appropriately weighted central exclusive $\chi_{c}$ event, we then generate the decay process \mbox{$\chi_c \to J/\psi \gamma\to \mu^+\mu^- \gamma$} through which exclusive $\chi_c$ production has been observed at the Tevatron. While the isotropic decay of the scalar \mbox{$\chi_{c0} \to J/\psi\gamma$} is trivial, the situation for $\chi_{c1}$ and $\chi_{c2}$, which have non-trivial polarization states that must be accounted for in the relevant decays, is not so simple. The calculation of the different decay distributions is, however, relatively straightforward and is outlined in full in Appendix \ref{decays}. In all cases, these decays have an integrated weight of unity, while we multiply the overall cross section by the relevant branching ratios, taken from~\cite{PDG}. For the $\chi_{c1}$, $\chi_{c2}$ and $J/\psi$ we generate the spin states in the helicity basis: that is, we generate the $\chi_c$($J/\psi$) polarization vectors in the $\chi_c$($J/\psi$) rest frame, before boosting along the spin quantization ($z$) axis and then rotating in the $z-{\bf p_{\chi(\psi)}}$ plane. Finally, to improve the overall efficiency we allow as input the option of specifying values for the experimentally significant cuts on the final state $\mu^+\mu^-$ pair, in particular the maximum pseudorapidity $|\eta|$ and the minimum $p_\perp$. All kinematic information for the produced particles is calculated and can be read out at the end of the run.
\section{Results}\label{res}
We begin, for the sake of comparison with the previous results of~\cite{Khoze04}, with a calculation of the $\chi_{c0}$ cross section in the forward limit. As in~\cite{Khoze04}, we use GRV94H0 partons~\cite{Gluck94} throughout\footnote{As already mentioned, the GRV94HO gluon PDF is very consistent with the more recent MSTW2008 and CTEQ LO and NLO gluon PDFs in the $0.05<x<1$ region.}, and we choose the value of $Q_\perp=0.85\,$ GeV as our infrared cut-off. Combining the perturbative and non-perturbative contributions, we find

\begin{equation}
\frac{{\rm d}\sigma_{\chi_{c0}}^{\rm approx}}{{\rm d}y_{\chi}}\bigg|_{y_\chi=0} = 80 \,{\rm nb} \; .
\end{equation}
Once we have corrected for the revised PDG value for the  total $\chi_{c0}$ width (which has decreased by a factor $\sim 1.4$), as well as the slightly revised value of the survival factor that we use, we find that this is in good agreement with the previous result quoted in~\cite{Khoze04}. Moreover, this is also in excellent agreement with the experimental value from the CDF collaboration~\cite{Aaltonen09}:
\begin{equation}
\frac{{\rm d}\sigma_{\chi_{c}}^{\rm exp}}{{\rm d}y_{\chi}}\bigg|_{y_\chi=0} = (76 \pm 14)\,{\rm nb} \; .
\end{equation}
We emphasise that this value was \emph{assumed}, rather than being observed, to correspond to $\chi_{c0}$ production, as the mass difference between the three $\chi_c$ states was not resolvable within the experimental set-up. 

\begin{figure}\begin{center}
\includegraphics[scale=0.95]{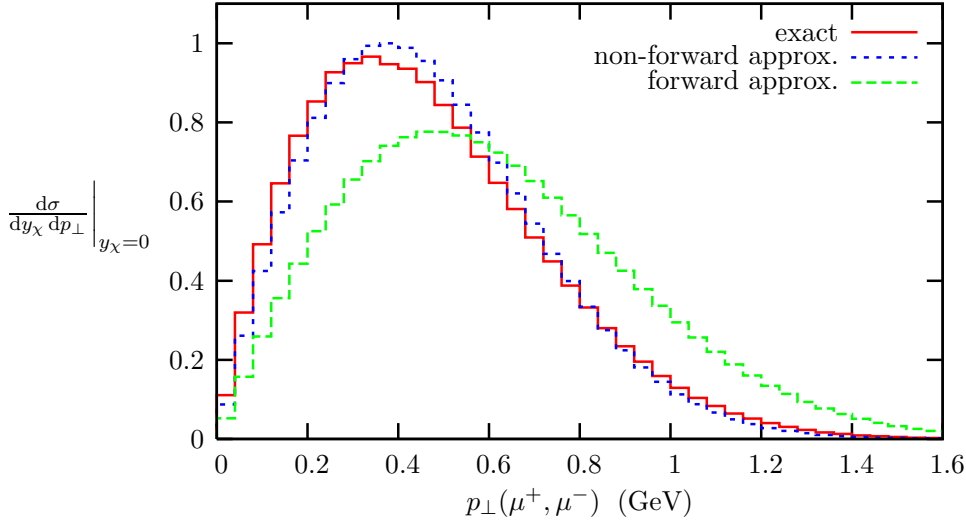}
\caption{Differential cross section (in arbitrary units) as a function of $\chi_{c0}$ $p_\perp$ as a result of exactly calculating the $p_\perp$ dependent $\chi_{c0}$ amplitude (`exact'), setting $p_\perp=0$ in the initial amplitude calculation (`forward approximation') and approximating the non-forward effects by an effective slope parameter $b_{\rm eff}$ (`non-forward approximation').} 
\label{chi0comp}
\end{center}\end{figure}

We now consider the effect of non-forward protons ($p_\perp \neq 0$) on this cross section (see also \cite{Kaidalov03,KKMRext}). By directly integrating over the $p_\perp$ dependent amplitude squared and fitting the resultant $p_{\chi_\perp}^2$ distribution according to (\ref{gfit0}), we find
\begin{equation}\label{beff0}
b^{\rm eff}_0 = 6.6 \,{\rm GeV}^{-2} \; ,
\end{equation}
that is, a steepening in the effective slope of the proton form factor\footnote{Recall that here $b^{\rm eff}/2$ is the slope of the `bare' amplitude {\it before} screening effects are included. The inclusion of absorptive effects (i.e. the gap survival factor) will further enlarge the $p_\perp$-slope of the experimentally observed forward proton distributions.}.
The effect of this is shown in Fig.~\ref{chi0comp}, where we have plotted the differential cross section as a function of $p_\perp(\chi_{c0})$ in the forward and non-forward limits discussed above. As a check, we plot both the exact and our fitted results for the non-forward limit: we can see that the match is sufficiently accurate for our purposes. As expected, the steeper slope corresponds to $p_\perp(\chi_{c0})$ being more sharply peaked at low $p_\perp$ values. This effect is enhanced by the prescription (\ref{minpres}) for the argument $Q_i^2$ of the skewed PDFs. In particular, we have introduced an infrared cut-off to avoid a contribution from the low $Q_\perp$ domain, where perturbation theory is not valid. However we can see that (\ref{minpres}) will have the average effect of pushing $Q_i$ into this `soft' region which we are neglecting, and therefore lower values of $p_\perp$ will be favoured, leading to a steepening of the effective slope. On the other hand, we could, instead of taking the minimum of the two gluon transverse momenta, take their average, which gives $b^{\rm eff}_0 \approx 5.6 \,{\rm GeV}^{-2}$. It is clear then that we expect some steepening in the effective slope, but the exact amount is very much tied up in the overall uncertainties of the calculation. Na\"{\i}vely, we might expect this to lead to a factor of $\sim$3 decrease in the production cross section, via the integration over the transverse momenta $p_{i_\perp}$, however we recall (see, for example,~\cite{Khoze04,epip,nns2}) that the total cross section depends on the ratio $S^2/b^2$, which depends only weakly on $b^2$; that is the reduction in the cross section caused by the increased slope is largely compensated by an increase in the survival factor for the more peripheral interaction, leading to only a small overall decrease in the $\chi_{c0}$ rate.
We next consider the $\chi_{c1}$ and $\chi_{c2}$ cross sections, which are calculated following the procedure outlined in Section \ref{MC}. Directly integrating over the $p_\perp$ dependent amplitude squared and fitting the resultant $p_{\chi_\perp}^2$ distribution according to (\ref{a1fit}) and (\ref{a2fit}) for $\chi_{c1}$ and $\chi_{c2}$ CEP, respectively, we find the following values for the effective slopes:
\begin{align}\label{beff1}
b_1^{\rm eff} = 4.6 \,{\rm GeV}^{-2}\; ,\\ \label{beff2}
b_2^{\rm eff} = 5.9 \,{\rm GeV}^{-2}\; ,
\end{align}
 We can see from Fig.~\ref{schipt} that while the $\chi_{c0}$ and $\chi_{c1}$ eikonal survival factors are approximately constant, the $\chi_{c2}$ eikonal survival factor has a strong dependence on the $p_\perp$ of the $\chi_{c2}$, which we read in from a grid in the Monte Carlo. Making use of (\ref{beff0}), (\ref{beff1}) and (\ref{beff2}) we find
\begin{equation}
S^2_{\rm eik}(\chi_0):S^2_{\rm eik}(\chi_1):\langle S^2_{\rm eik}(\chi_2)\rangle \approx 0.098:0.15:0.22 \; .
\end{equation}
We emphasise that the gap survival factor $S^2$ depends on the effective slope $b^{\rm eff}$ of the {\it bare} (non-screened) CEP amplitude which, in turn, depends explicitly on the $p_\perp$ of the outgoing protons, see \cite{KKMRext}. In particular, at $b=4$ GeV$^{-2}$ we obtain $S^2_{\rm eik}=0.046$, while in the case of CEP of the $\chi_{c0}$, where $b_{\rm eff}^0 = 6.6$ GeV$^{-2}$, we obtain the value $S^2_{\rm eik}=0.098$, i.e. a factor of two larger. As discussed above, such an increase in $S^2$ largely compensates the decrease in the CEP cross section caused by a smaller phase space  in $p_\perp$ occupied by the final state protons (due to a larger $b^{\rm eff}$). Recalling that the CEP event rate depends on the ratio $S^2/b^2$ (rather than on $S^2$) \cite{Khoze04}, this can serve as a warning regarding the conclusions made in Refs. \cite{strik,GLMM} about the CEP rates based solely on the evaluations of $S^2$ for  protons with $p_\perp = 0$. In addition, we have already shown in Section 3 that neglecting the $p_\perp \neq 0$ effects in the structure of the hard production subprocess can lead to a significant underestimate in the production rate of states with $J^{\rm P}$ other than $0^+$. 
\begin{figure*}\begin{center}
\includegraphics[scale=0.95]{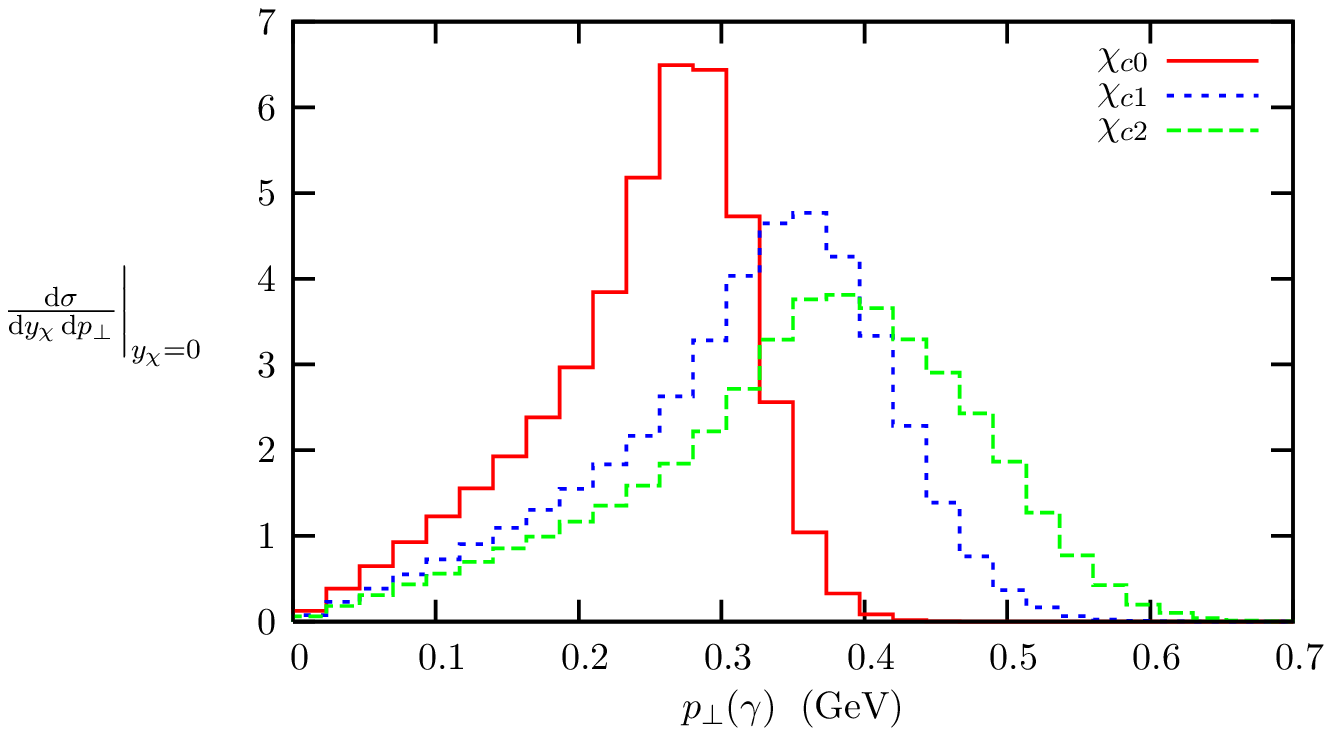}
\includegraphics[scale=0.95]{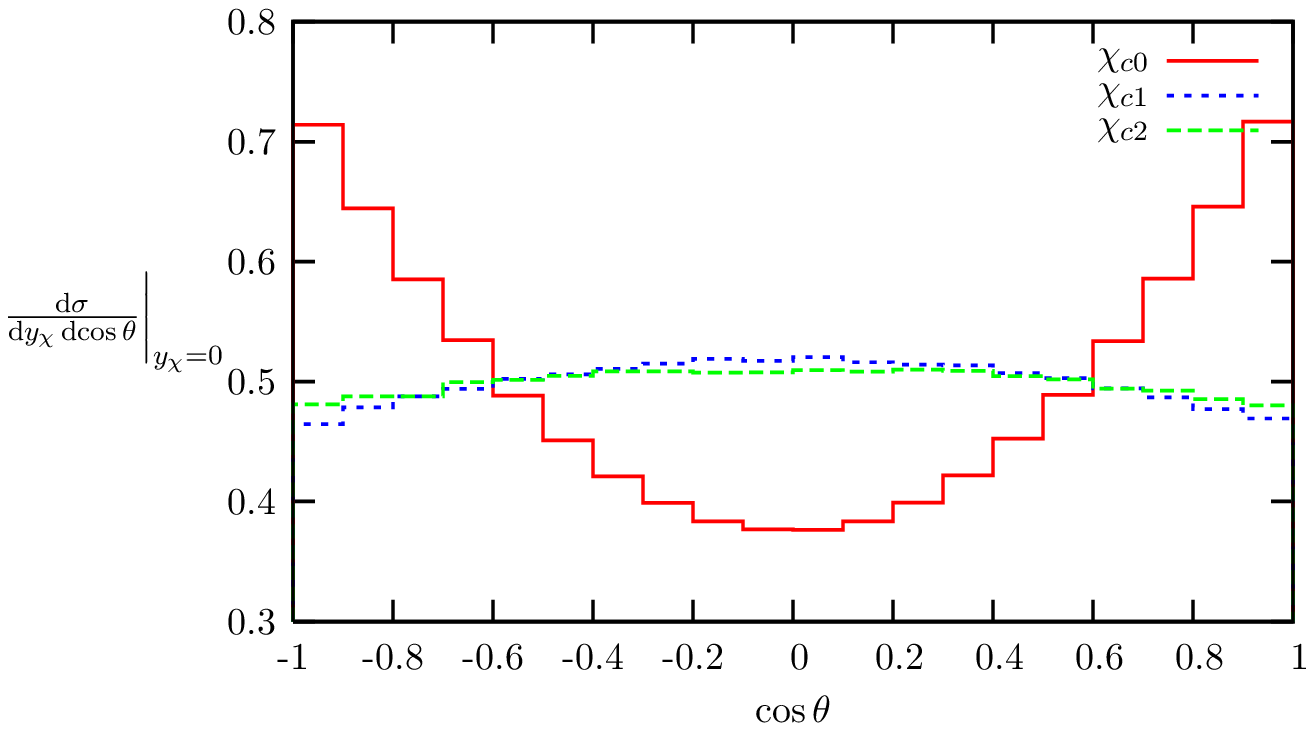}
\caption{$\chi_{c0}$, $\chi_{c1}$ and $\chi_{c2}$ differential cross sections, without experimental cuts, as a function of the photon $p_\perp$ and $\cos\theta$, where $\theta$ is the angle between the $\mu^+$ momentum in the $J/\psi$ rest frame and the direction of the Lorentz boost from the $J/\psi$ rest frame to the lab frame. Integrated cross sections normalised to unity before cuts are imposed in all cases.} \label{angplot}
\end{center}\end{figure*}
Returning to the calculation of the higher spin $\chi_c$ cross sections we find, by including the relevant branching ratios (evaluated at rapidity $y_\chi=0$ in all cases),
\begin{align}\label{rates}
\frac{\Gamma^{\chi_0}_{J/\psi\gamma}}{\Gamma^{\chi_0}_{\rm tot}}\frac{{\rm d}\sigma_{\chi_{c0}}^{\rm pert}}{{\rm d}y_{\chi}}:\frac{\Gamma^{\chi_1}_{J/\psi\gamma}}{\Gamma^{\chi_1}_{\rm tot}}\frac{{\rm d}\sigma_{\chi_{c1}}^{\rm pert}}{{\rm d}y_{\chi}}:\frac{\Gamma^{\chi_2}_{J/\psi\gamma}}{\Gamma^{\chi_2}_{\rm tot}}\frac{{\rm d}\sigma_{\chi_{c2}}^{\rm pert}}{{\rm d}y_{\chi}}
\!\approx\! 1: 0.8:0.6 \; .
\end{align}
Thus, within the perturbative framework, the expected contributions of the three $\chi_c$ states to the Tevatron data are of comparable size, despite the initial suppression in the $\chi_{c1}$ and $\chi_{c2}$ production amplitudes. The previous assumption that the Tevatron events correspond to purely $\chi_{c0}$ CEP may therefore be unjustified. To give a prediction for the total cross section we must take into account the so-called `enhanced' absorptive effects, which break the soft-hard factorization previously assumed in the evaluation of the $\chi_{c0}$ CEP cross section in~\cite{Khoze04}. The generalisation of the simple two channel eikonal model to include these enhanced rescattering effects is outlined in~\cite{nns2}, where the effect of including both eikonal and enhanced screening corrections, as well as non-forward outgoing protons in the hard matrix element, can be roughly accounted for by the introduction of an `effective' survival factor $S^2_{\rm eff}$. Recalling (\ref{s:ineq}), it is found that the effect of enhanced absorption
 for the CEP of the light $\chi_c$ is quite strong (see also \cite{bbkm,strik,GLMM}). In particular, ignoring for simplicity any impact parameter ${\bf b}$ dependence, we have
\begin{figure*}\begin{center}
\includegraphics[scale=0.95]{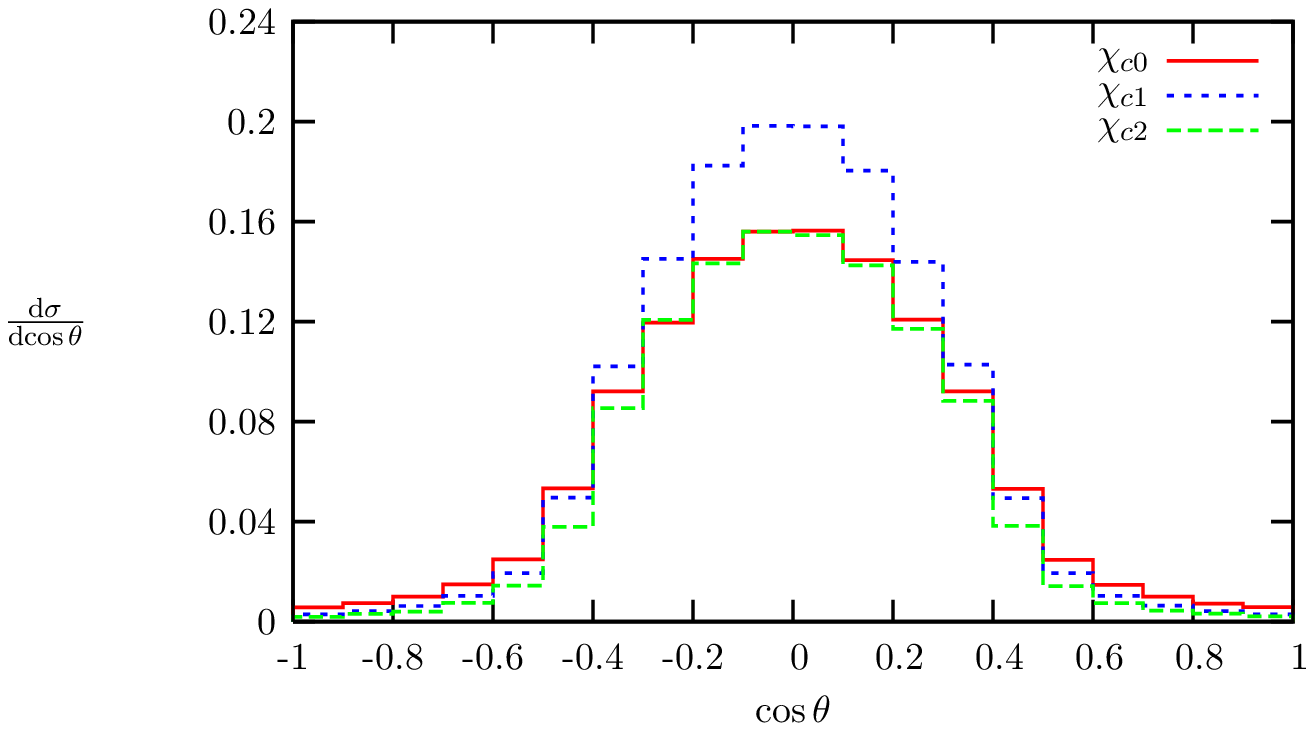}
\includegraphics[scale=0.95]{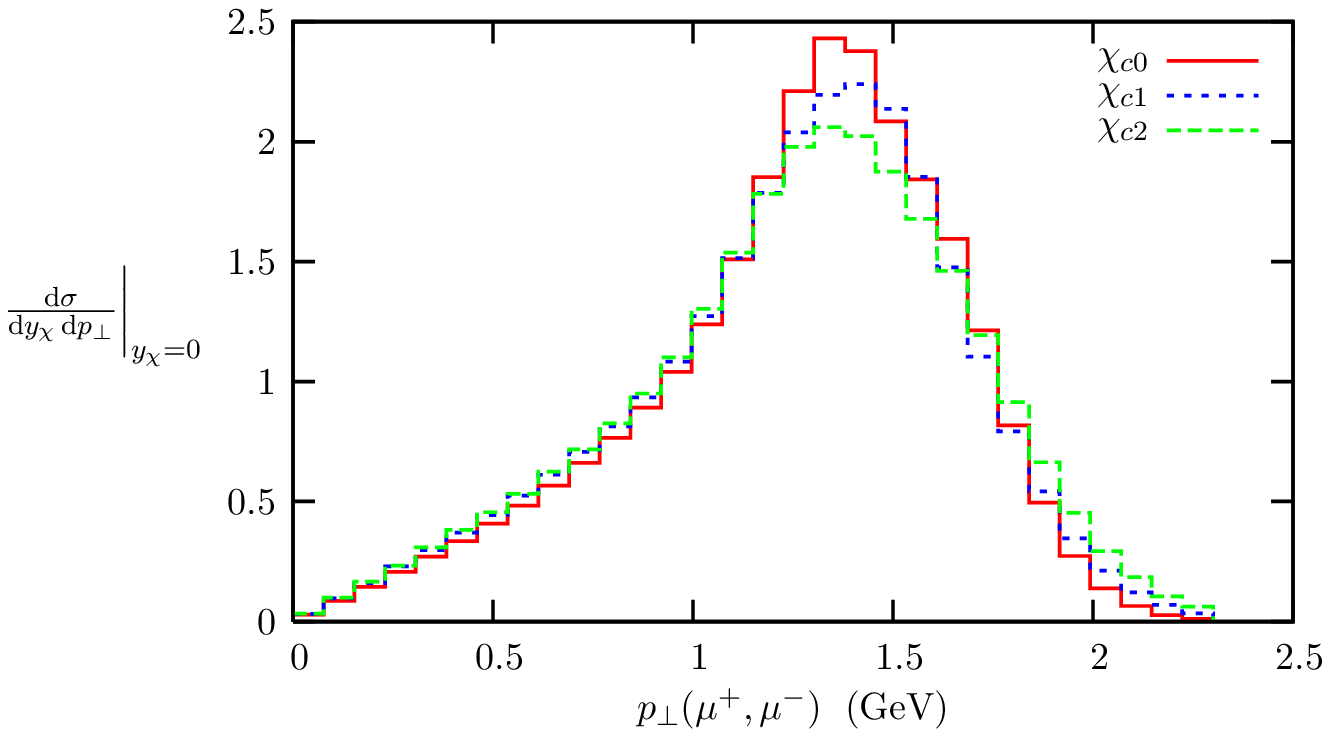}
\caption{Differential cross sections as a function of $\cos\theta$, with experimental cuts, and the $\mu^+, \mu^-$ $p_\perp$, without experimental cuts. Integrated cross sections normalised to unity before cuts are imposed in all cases.} \label{angplotcut}
\end{center}\end{figure*}
\begin{equation}
\langle S^2_{\rm eff} \rangle \approx \langle S^2_{\rm enh}\rangle \times \langle S^2_{\rm eik}\rangle \approx \frac{1}{3}\, \langle S^2_{\rm eik}\rangle \; ,
\end{equation}
with the enhanced survival factor $S^2_{\rm enh}$ (at our present level of understanding) approximately the same for all three $\chi_c$ states. Multiplying by this suppression factor and including the contribution from the three $J$ states to the observed cross section, we can then use the branching ratio Br($\chi_{c0} \to J/\psi + \gamma$), as was done for the CDF data, to produce a rough value for the `$\chi_{c0}$' cross section at Tevatron energies (prior to any corrections due to the varying experimental acceptances of the $\chi_c$ states),
\begin{equation}
\frac{{\rm d}\sigma_{\chi_{c}}^{\rm tot}}{{\rm d}y_{\chi}}\bigg|_{y_\chi=0}\approx 65 \,{\rm nb} \; .
\end{equation}
Here we stress that a sizeable proportion of the observed events are predicted to correspond to $\chi_{c1}$ and $\chi_{c2}$ CEP. Thus, the combination of including the more general enhanced rescattering effects, which leads to a reduction in the predicted rate, and the contribution of the higher spin $\chi_{c1}$ and $\chi_{c2}$ states, which leads to an increase in the predicted rate, leaves the perturbative prediction for the total $\chi_c$ cross section at the Tevatron largely unchanged and, crucially, still in good agreement with the experimental data.

At this point we need also to consider the non-perturbative contribution to the cross section. Leaving the explicit evaluation for future work, we simply note that the previous calculations of ~\cite{Peng95,Stein93} show that the non-perturbative contributions of the three $\chi_c$ states are also comparable, with the general results of Section \ref{Reggesec} suggesting that, at least for the $\chi_{c0}$ and $\chi_{c1}$, they will have a similar $p_\perp$ dependence to that outlined in Section \ref{MC}. We can therefore reasonably assume that the relative values of (\ref{rates}) are approximately correct. However, given the overall uncertainty in the perturbative and non-perturbative cross section calculations, we note that the precise ratio of the $\chi_c$ cross sections cannot be stated with certainty at this time. While our results suggest that some fraction of the 65 $\pm$ 10 candidate `$\chi_{c0}$' events observed at CDF are in fact $\chi_{c1}$ or $\chi_{c2}$ events, we can make no definitive prediction for their precise relative contributions.

In fact, we have noted that the perturbative contribution to $\chi_{c2}$ CEP is strongly suppressed due to the $J_z=0$ selection rule, for which it decouples from two real gluons. However, for the non-perturbative contribution we have no such selection
rule, and we therefore cannot exclude the possibility that the $\chi_{c2}$ non-perturbative contribution is
dominant. The above-mentioned Pomeron models suggest that this may be the case, although there remains a large degree of uncertainty in how to perform these calculations, and in particular which model of the Pomeron to choose.

Putting aside the question of normalisation, we might hope to be able to distinguish between the three states by studying the angular and kinematical distributions of the final state particles as modelled in SuperCHIC, which should not depend strongly on the overall production rate. In Fig.~\ref{angplot} we show the polar angular distributions of the $\mu^+ \mu^-$ pair and the $p_\perp$ distributions of the photon for the three $\chi$ states, as given by SuperCHIC. For the $p_\perp$ distribution we can see the clear separation in the $\chi$ masses coming from the position of the Jacobian peaks as $p_\perp$ approaches the photon energy $E_\gamma \sim M_\chi - M_\psi$, although in the current experimental set-up we know it is not possible to resolve this separation. 

The angular distribution is more interesting: there is clearly a significant difference between the $\chi_{c0}$ and the $\chi_{c1}$ and $\chi_{c2}$ cases with, as expected from helicity conservation, the $\chi_{c0}$ decaying into purely transversely polarized $J/\psi$'s, while this is not the case for the $\chi_{c1}$ and $\chi_{c2}$. This in principle provides a way to determine if $\chi_{c1}$ and $\chi_{c2}$ mesons are being produced, irrespective of the particular mass resolution of the experiment. Unfortunately, this does not appear to be the case in practice, as we have yet to include the experimental cuts on the kinematics of the final state particles. At CDF we require in particular that the muon pseudorapidity $|\eta|<0.6$ and the muon $p_\perp>1.4\,$~GeV. In Fig.~\ref{angplotcut} we show the angular distribution as before but with these cuts introduced, and immediately we can see that the clear difference in shape has not survived. We also show the $p_\perp$ distribution of the muons, from which it is clear that a sizeable fraction of the events will not pass the cuts (recalling that the cut must be passed by both muons), and the pseudorapidity cut further enhances this effect. Moreover, in the low (high) $\theta$ region the $\mu^{+(-)}$ will be directed along the motion of the $J/\psi$ with high $p_\perp$, while the $\mu^{-(+)}$ will be directed against the motion of the $J/\psi$ and will therefore have low $p_\perp$. This is clear from Fig.~\ref{angplotcut} where the $\cos\theta \approx +1,-1$ events which would have allowed us to distinguish between $\chi_{c1,2}$ and $\chi_{c0}$ production have not been accepted. We note that other potentially interesting variables, such as the difference in the azimuthal angle of the muons $\Delta\phi_{\mu\mu}$, appear to be equally unpromising. It therefore seems that it will be very hard, given the experimental set up and the low statistics available at the present time, to distinguish between the three $\chi_c$ states via the experimentally considered decay chain, although with more detailed analysis and/or higher statistics this conclusion may change.
Another potential solution to this issue could be to consider a different decay chain, for example the direct decay of the $\chi_c$ to charged hadrons, etc. (see Section~6 below).

Finally, we can also see that the $\chi_{c1}$ has a slightly  higher acceptance, with in particular (at $y_\chi = 0$)
\begin{equation}
\frac{{\rm d}\sigma^{\rm cuts}_{\chi_{c0}}}{{\rm d}\sigma^{\rm tot}}:\frac{{\rm d}\sigma^{\rm cuts}_{\chi_{c1}}}{{\rm d}\sigma^{\rm tot}}:\frac{{\rm d}\sigma^{\rm cuts}_{\chi_{c2}}}{{\rm d}\sigma^{\rm tot}}\approx 16\%:17\%:15\%    \; .
\end{equation}
The acceptance is therefore reasonably uniform, and so should not present a significant obstacle when considering $\chi_{c1}$ and $\chi_{c2}$ production experimentally.
\section{Summary and Outlook}\label{outlook}

Motivated by the recent experimental observation of exclusive $\chi_{c}$ events at the Tevatron, we have updated the earlier studies of central exclusive scalar $\chi_{c0}$ meson production to include $\chi_{c1}$ and $\chi_{c2}$ mesons. 
Due to the low scale, $M_{\chi_c}/2$, and  very large rapidity gap coverage ($\Delta\eta \simeq7.4$ units) in the CDF measurement \cite{Aaltonen09}, the contamination from processes in which the incoming protons dissociate is relatively small.
The CDF $\chi_{c}$ event selection therefore effectively ensures that they come from the exclusive reaction,
 $p\bar{p}\to p\ +\ \chi_c\ +\bar{p}$. 
 
Although $\chi_{c0}$ production was previously assumed to be dominant, we find that the $\chi_{c0}$, $\chi_{c1}$ and $\chi_{c2}$  rates for the experimentally considered $\chi_c \to J/\psi \gamma \to \mu^+ \mu^-\gamma$ decay process are in fact comparable. We have developed a new Monte Carlo event generator, SuperCHIC, which models the central exclusive production of the three $\chi_c$ states via this decay chain, and have used this to explore possible ways of distinguishing them, given that their mass differences are not resolvable within the current experimental set-up. Although we find that the severity of current experimental cuts appears to preclude this discrimination, 
the acceptance does not change crucially between the three states and so our conclusions regarding the overall rates remain unchanged. This therefore raises the interesting possibility that exclusive $\chi_{c1}$ and $\chi_{c2}$ production has already been observed at the Tevatron. 
Higher statistics and/or a broader acceptance coverage for the photon and leptons could help discriminate between the $\chi_{c}$ mesons via differences in the angular correlations between the final-state particles.
We note also that the addition of forward proton detectors would certainly allow discrimination between the different $C$-even states via the measurement of the relative azimuthal angular distribution between
the outgoing protons. 

To further resolve the spin-parity assignment issue in the absence of forward proton detectors, it would be instructive to observe central exclusive $\chi_c$ production in other decay channels, in particular
$\pi\pi$ or $K\bar{K}$, see \cite{Khoze04}. These modes are ideally suited for spin-parity analysis: the $\pi\pi$ or $K\bar{K}$ decay modes of the $\chi_{c0}$ meson have a branching
fraction of about 1$\%$, while these decay channels are forbidden for $\chi_{c1}$ and suppressed by about a factor of 5 for the $\chi_{c2}$ relative to the $\chi_{c0}$, in contrast to the  
$\chi_{c2}$ relative enhancement for the $J/\psi\gamma$ channel. Another interesting mode for discriminating between the CEP of different $\chi_c$ states is $\chi_c\to p\bar{p}$, since the branching fraction
for $\chi_{c0}$ ($\simeq 0.024\%$) is a factor of 3 higher than that for  $\chi_{c1,2}$ \cite{PDG}. The $\Lambda\bar{\Lambda}$ mode (branching fraction for $\chi_{c0} \simeq 0.034\%$)
could also be important for spin-parity analyzing.\footnote{Rough estimates show that the background from continuum $\pi\pi$, $K\bar{K}$, $p\bar{p}$ and $\Lambda\bar{\Lambda}$ central production
should be quite manageable. This is in accord with measurements
in two-photon collisions \cite{2g}, where the $\chi_{c0,2}$ resonances decaying to  $\pi\pi$ and $K\bar{K}$ final states are clearly seen.}

In the case of two-body final states ($\pi\pi$, $KK$, $p\bar{p}$) where the very forward protons are not detected, further cuts can be imposed to reduce the contribution of events where the protons dissociate (via single and double diffractive dissociation). These include, for example, cuts  on the transverse momentum of the resonance and 
 on the final particles' accoplanarity angle (in the frame where the rapidity
 of the resonance is zero).\footnote{This procedure is similar to that used in the separation of exclusive  lepton-pair production via photon-photon fusion, $pp \to p + l^ + l^- + p$, see~\cite{KMROlum} for more details.}

In this paper we have focused on $\chi_{c}$ meson production at the Tevatron. It is of course straightforward to extend our results
to the LHC, and we will consider this is in a future study~\cite{Langetal}. Note that we do not expect the $\chi_c$ CEP rate to have a strong energy dependence
when going from the Tevatron to the LHC. The growth of the bare amplitude caused by the increase in the gluon
density at smaller $x$ is compensated by a smaller gap survival factor at the larger LHC energies,
especially $S^2_{\rm enh}$, the value of which decreases due to the larger rapidity interval available for the `enhanced' absorptive corrections, see Fig.~1(d).
Indeed, the measurement of the ratios of the CEP rates
at the two different (Tevatron and LHC) collider energies could allow the effects of  enhanced absorption to be probed, since in these cross section ratios
various uncertainties (for example, NLO corrections to the $gg\to\chi$ transition etc.)
would cancel out.

The issue of forward proton detection is more relevant at the LHC
(in particular for $\chi_b$ states) as
the planned near-beam proton detectors, see Refs.~\cite{AR,FP420,RP}, would allow us to measure the outgoing very forward protons. As we have already
noted, the azimuthal angle distributions would provide interesting additional information with which we could discriminate between
different $J^{PC}$ states as well as investigate the dynamics of the survival factors $S^2$. Moreover,
as pointed out in \cite{KMRforw}, measuring the transverse momentum and azimuthal angle correlations between the outgoing protons
would allow us to probe the proton opacity $\Omega(s,b_t)$ and perform a detailed test of the {\it whole}
diffractive formalism. Even before the forward proton detectors become operational, a broad programme of heavy quarkonium studies can be performed with
the existing LHC detectors (ALICE, ATLAS, CMS and LHCb), especially if the rapidity gap coverage is increased
by using  forward shower counters (FSC) along the beam line \cite{FSC}, thereby allowing detection and triggering on rapidity
gaps in diffractive events.

In future work we will also consider $\chi_{b}$ and $\eta_{b,c}$ meson CEP and include them in the Monte Carlo. The perturbative $\chi_{b0}$ CEP rate was calculated in~\cite{Khoze00a,Khoze04} analogously to the $\chi_{c0}$. 
While the higher $M_\chi$ scale results in a more perturbatively reliable prediction, the reduced rate suggests that exclusive $\chi_b$ production, observed through the $\chi_{b} \to \Upsilon(1{\rm S})\gamma \to  \mu^+ \mu^-\gamma$ decay chain, may only be relevant at the LHC. The experimental values for the $\chi_{bJ} \to \Upsilon(1{\rm S})\gamma$ branching ratios carry larger uncertainties, but a similar hierarchy to the $\chi_c$ case is observed, and so the contribution of higher spin states should once again be considered, with in particular the result of (\ref{comp}) for the relative rates remaining valid. However, we note that the higher mass scale will result in a stronger suppression of the higher spin states relative to the $\chi_{b0}$ than in the $\chi_c$ case. In particular, for $\chi_{b1}$ production we have an explicit factor of $M^2_\chi$ in the denominator of (\ref{comp}), while the larger expected value of $\langle Q^2_\perp \rangle$ will result in a stronger suppression of the $\chi_{b2}$ state. 

The $\eta_{c,b}$ CEP cross sections can be calculated by using the same formalism as for the $\chi$ mesons, the only difference being that for the $L=0$ quarkonium state 
the $gg\to\eta$ vertex is proportional to the value of the wave function $\phi(0)$ at the origin and not to $\phi'(0)$ as in (\ref{v0norm}). The vertex $V_\eta$ should therefore be normalised to the leptonic width of the $J/\psi$ (or $\Upsilon$ for $\eta_b$) decay -- the vector mesons from the same $L=0$ multiplet. Note also that for the heavier $\chi_b$ or $\eta_b$ mesons we expect a slightly lower slope $b^{\rm eff}$ (due to a larger mean $\langle Q^2_\perp \rangle$) and a larger value of $S^2_{\rm enh}$ (due to a smaller rapidity
interval available for the `enhanced' absorptive corrections, Fig.~1(d)). Preliminary estimates indicate that the $\eta_c$ CEP rate (which we recall from~\cite{Kaidalov03} will be proportional to $\langle { p}_{1_\perp}^2 { p}_{2_\perp}^2 \rangle/ \langle { Q}_\perp^2 \rangle^2$) is expected to be about two orders
of magnitude lower than in the $\chi_{c0}$ case.
 
We are also planning to revisit $\gamma\gamma$ CEP
in a wider interval of photon $E_T$, rapidity and di-photon mass $M$
than that considered in \cite{KMRSgg} and to include this in the Monte
Carlo generator.
Note that the measurement of the ratio of $\gamma\gamma$ CEP
at $E_T$ = 5 GeV to that of $\chi_b$ production may allow us to reduce
various uncertainties in the calculations, with in particular the dependence on the survival
factors cancelling out.

Finally, we note that the spin-parity analyzing  properties of central exclusive production could shed light on the dynamics of the {\it zoology} of `exotic'
charmonium-like states ($X,Y,Z$) which have been discovered in the last few years (see for example \cite{PDG,cc}), and whose nature and in many cases spin-parity assignment still remain unclear.\\

\section*{Acknowledgements}

We thank Albert De Roeck, Aliosha Kaidalov, Alan Martin, Risto Orava, Jim Pinfold
Rainer Schicker,
  Oleg Teryaev, and especially  Mike Albrow
for useful and encouraging discussions.
MGR, LHL and WJS thank the IPPP at the University of Durham for hospitality.
The work was supported by  grant RFBR
07-02-00023, by the Russian State grant RSGSS-3628.2008.2. LHL acknowledges financial support from the University of Cambridge Domestic Research Studentship scheme.

\appendix
\renewcommand{\theequation}{A.\arabic{equation}}

\section{$\chi_c \to gg$ amplitudes}\label{chis}

\subsection{$\chi_{c0}$}
We use the formalism of \cite{Kuhn79} and the kinematics of Section \ref{theory} throughout. The general colour-averaged vertex for the coupling to two gluons has the form
\begin{equation}\label{AV0}
V_0\equiv\epsilon_1^\alpha\epsilon_2^\beta V^0_{\alpha\beta}=\sqrt{{\frac{1}{6}}}\frac{c}{M}(I^0_1(M^2+(q_1q_2))-2I^0_2) \; ,
\end{equation}
where $\epsilon_{1,2}$ are the polarization vectors of the incoming gluons and
\begin{align}
q_1 &= x_1p_1+Q_\perp - p_{1_\perp}\; ,\\
q_2 &= x_2p_2-Q_\perp - p_{2_\perp}\; ,
\end{align}
while we define
\begin{align}
I_1^0&=F_{\mu\nu}^1F^{2,\mu\nu} \; ,\\
I_2^0&=q_1^\nu F_{\mu\nu}F^{2,\mu\sigma}q_{2,\sigma} \; .
\end{align}
Here $F_{\mu\nu}$ is the usual field strength tensor for the gluons and
\begin{equation}\label{c}
c=\frac{1}{2\sqrt{N_C}}\frac{4g_s^2}{(q_1q_2)^2}\sqrt{\frac{6}{4\pi M}}\phi_c'(0) \; ,
\end{equation}
where $g_s$ is the strong coupling and $\phi_c'(0)$ is the derivative of the $\chi_c$ radial wavefunction at the origin. Note that our definition of $c$ differs from that of Ref.~\cite{Kuhn79} by a factor
\begin{equation}
\left\langle 3i;\overline{3}k|1\right\rangle t^{a}_{ij}t^b_{jk}= \frac{\delta^{ab}}{2\sqrt{N_C}}\to \frac{1}{2\sqrt{N_C}} \; ,
\end{equation}
where $\left\langle 3i;\overline{3}k|1\right\rangle$ is the colour space Clebsch-Gordon coefficient for the colour singlet quark configuration, and we have averaged over the gluon colour indices $a,b$ in the last step. Recalling (\ref{Vnorm}), we make the replacement  $\epsilon_1^\mu \epsilon_2^\nu V_{\mu\nu}\to \frac{2}{s}p_1^\mu p_2^\nu V_{\mu\nu}$ and use the gauge invariance of $V_{\mu\nu}$ to give
\begin{align}
I_1^0&=2(q_{1_{\perp}}q_{2_{\perp}}) \; ,\\
I_2^0&=q_{1_{\perp}}^2q_{2_{\perp}}^2 \; .
\end{align}
Here we have made use of the identity
\begin{equation}
(q_1q_2)=\frac{1}{2}(M^2-q_{1_{\perp}}^2-q_{2_{\perp}}^2) \; .
\end{equation}
We therefore obtain
\begin{equation}\label{v0}
V_0=\sqrt{\frac{1}{6}}\frac{c}{M_\chi}((q_{1_{\perp}}q_{2_{\perp}})(3M_\chi^2-q_{1_{\perp}}^2-q_{2_{\perp}}^2)-2q_{1_{\perp}}^2q_{2_{\perp}}^2)  \; 
\end{equation}
\subsection{$\chi_{c1}$}\label{chis1}
The general colour-averaged vertex has the form
\begin{equation}
V_1=-\frac{ic}{2}(I_1^1+I_2^1) \; ,
\end{equation}
where
\begin{align}
I_1^1&=\epsilon^{\mu\nu\alpha\beta}\epsilon^{\chi*}_\beta F^1_{\mu\nu}F^2_{\alpha\gamma}q_2^\gamma \; \\
I_1^2&=\epsilon^{\mu\nu\alpha\beta}\epsilon^{\chi*}_\beta F^2_{\mu\nu}F^1_{\alpha\gamma}q_1^\gamma \; .
\end{align}
Here $\epsilon^\chi_\beta$ is the $\chi_1$ polarization vector and $\epsilon^{\mu\nu\alpha\beta}$ is the antisymmetric Levi-Civita tensor. This gives
\begin{equation}
V_1=-\frac{2ic}{s}\epsilon^{\mu\nu\alpha\beta}\epsilon^{\chi*}_\beta p_{1,\nu}p_{2,\alpha}((q_{2\perp})_\mu(q_{1\perp})^2-(q_{1\perp})_\mu(q_{2\perp})^2) \; .
\end{equation}
\subsection{$\chi_{c2}$}\label{chis2}
The general colour-averaged vertex has the form
\begin{equation}
V_2=-c\sqrt{2}MI_2^2 \; ,
\end{equation}
where
\begin{equation}
I_2^2=\epsilon^{*\mu\alpha}_\chi F^{1\beta}_\mu F^2_{\alpha\beta} \; .
\end{equation}
$\epsilon^{\mu\alpha}_\chi$ is the $\chi_2$ polarization tensor, which satisfies
\begin{align}\label{eps2}
\epsilon_{\mu\nu}&=\epsilon_{\nu\mu}\; , \qquad \epsilon^{\phantom{\mu}\mu}_\mu=0 \; ,
\qquad \epsilon_{\mu\nu}P^\mu_\chi=0 \; ,\\ \label{eps2pol}
\sum_{pol}\epsilon_{\mu\nu}\epsilon^*_{\alpha\beta}&=\frac{1}{2}(\mathbb{P}_{\mu\alpha}\mathbb{P}_{\nu\beta}+\mathbb{P}_{\mu\beta}\mathbb{P}_{\nu\alpha})
-\frac{1}{3}\mathbb{P}_{\mu\nu}\mathbb{P}_{\alpha\beta}\; ,\\
\mathbb{P}^{\mu\nu}&\equiv-g^{\mu\nu}+\frac{P^\mu_\chi P^\nu_\chi}{M^2} \; .
\end{align}
We therefore obtain
\begin{equation}\label{v2}
V_2=\frac{\sqrt{2}cM}{s}(s(q_{1\perp})_\mu(q_{2\perp})_\alpha+2(q_{1\perp}q_{2\perp})p_{1\mu}p_{2\alpha})\epsilon_\chi^{*\mu\alpha} \; .
\end{equation}
\renewcommand{\theequation}{B.\arabic{equation}}
\section{$\chi_c$ and $J/\psi$ decay amplitudes}\label{decays}
We will make use of the identities
\begin{align}\label{id1}
\epsilon^{\mu\nu\alpha\beta}\epsilon_{\tilde{\mu}\tilde{\nu}\alpha\beta}&=-4\,\delta^{[\mu}_{\tilde{\mu}}\,\delta^{\nu]}_{\tilde{\nu}}  \; ,\\ \label{id2}
\epsilon^{\mu\nu\alpha\beta}\epsilon_{\tilde{\mu}\tilde{\nu}\tilde{\alpha}\beta}&=-3\mathcal{!}\,\delta^{[\mu}_{\tilde{\mu}}\, \delta^\nu_{\tilde{\nu}}\, \delta^{\alpha]}_{\tilde{\alpha}}  \; ,\\ \label{id3}
\epsilon^{\mu\nu\alpha\beta}\epsilon_{\tilde{\mu}\tilde{\nu}\tilde{\alpha}\tilde{\beta}}&=-4\mathcal{!}\,\delta^{[\mu}_{\tilde{\mu}}\, \delta^\nu_{\tilde{\nu}}\, \delta^{\alpha}_{\tilde{\alpha}}\,\delta^{\beta]}_{\tilde{\beta}} \; .
\end{align}
\subsection{$\chi_c(0^{++}) \to J/\psi + \gamma$}\label{chi0decay}
The scalar $\chi_0$ decays into a transversely polarized photon with a uniform angular distribution in its rest frame, and conservation of angular momentum therefore requires the $J/\psi$ to be transversely polarized.
\subsection{$\chi_c(1^{++}) \to J/\psi + \gamma$}\label{chi1decay}
The amplitude that is expected to dominate (as it is the amplitude which corresponds to the dipole transition~\cite{Kuhn79}) is of the form
\begin{equation}
A_1 \sim \epsilon^{\mu\nu\alpha\beta} \epsilon^\chi_\mu \epsilon^{\psi*}_\nu p^{\gamma}_\alpha \epsilon_\beta^{\gamma*} \; .
\end{equation} 
Squaring and summing over photon polarizations, and making use of (\ref{id2}), we find
\begin{equation}
\sum_{\epsilon_\gamma} |A_1|^2 \!\sim\! |(p_\gamma \epsilon_\psi)|^2 + |(p_\gamma \epsilon_\chi)|^2 + 2\,{\rm Re}[(\epsilon_\chi \epsilon_\psi)(\epsilon_\chi^* p_\gamma)
(p_\gamma \epsilon^{*}_\psi)] \; .
\end{equation}
The normalisation is given by summing over $J/\psi$ polarizations and making use of (\ref{id1}) and (\ref{id3})
\begin{equation}
|A^{{\rm norm}}_1|^2 \!\sim\! |(\epsilon_\chi p_\gamma)|^2 + \frac{(p_\gamma p_\psi)}{M_\psi^2}( (p_\gamma p_\psi) + 2\, {\rm Re}[(\epsilon_\chi^* p_\gamma)(\epsilon_\chi p_\psi)]) \; .
\end{equation}
We then divide by the normalisation factor to give the relative amplitudes squared for the three different $\chi_1$ polarizations.
\subsection{$\chi_c(2^{++}) \to J/\psi + \gamma$}\label{chi2decay}
Following similar arguments to the $\chi_1$ case (that is, assuming the dipole transition dominates), we can write the invariant amplitude as
\begin{equation}\label{A2}
A_2=\epsilon^{\mu\alpha}_\chi (F^\gamma)^\beta_{\phantom{\beta}\mu} (F^\psi)_{\alpha\beta} \; .
\end{equation}
To study angular correlations it is necessary to consider the explicit form of the $\chi_2$ polarization tensor. This represents the irreducible tensor operator for $J=2$ angular momentum, which can be decomposed in terms of
the spin and orbital polarization vectors~\cite{Kuhn79}
\begin{equation}
\epsilon^{(J_{_Z})}_{\mu\nu}=\sum_{S_{_Z},m} \epsilon^{(S_{_Z})}_\mu \epsilon^{(m)}_\nu \left\langle S=1,L=1,S_{_Z},m|J=2,J_{_Z}\right\rangle \; ,
\end{equation}
where $\left\langle S=1,L=1,S_{_Z},m|J=2,J_{_Z}\right\rangle$ are the Clebsch-Gordon coefficients and $\epsilon^{(S_{_Z})}_\mu, \epsilon^{(m)}_\nu$ have the usual explicit representation in (say) the $\chi_2$ rest frame. We can thus decompose the 5 polarization states as
\begin{align}
\epsilon^{_{+2}}_{\mu\nu}&=\epsilon^{_+}_\mu \epsilon^{_+}_\nu \; ,\\
\epsilon^{_{+1}}_{\mu\nu}&=\sqrt{\frac{1}{2}}(\epsilon^{_+}_\mu \epsilon^{_0}_\nu + \epsilon^{_0}_\mu \epsilon^{_+}_\nu) \; ,\\
\epsilon^{_0}_{\mu\nu}&=\sqrt{\frac{1}{6}}(\epsilon^{_+}_\mu \epsilon^{_-}_\nu + 2\,\epsilon^{_0}_\mu \epsilon^{_0}_\nu +\epsilon^{_-}_\mu \epsilon^{_+}_\nu) \; ,\\
\epsilon^{_{-1}}_{\mu\nu}&=\sqrt{\frac{1}{2}}(\epsilon^{_-}_\mu \epsilon^{_0}_\nu + \epsilon^{_0}_\mu \epsilon^{_-}_\nu) \; ,\\
\epsilon^{_{-2}}_{\mu\nu}&=\epsilon^{_-}_\mu \epsilon^{_-}_\nu \; .
\end{align}
Returning to (\ref{A2}), we obtain
\begin{align}\nonumber
\sum_{\epsilon_\gamma} |A_2|^2 &\sim 2\, {\rm Re}[\epsilon^{\mu\alpha} \epsilon^{*\nu\sigma} p^\gamma_\mu ( (p_\gamma p_\psi) \epsilon^{\psi}_\sigma-
(p_\gamma \epsilon_\psi)p^\psi_\sigma )(\epsilon^{*\psi}_\alpha p^\psi_\nu-\epsilon^{*\psi}_\nu p^\psi_\alpha)] \\ \nonumber
&-\epsilon^{\mu\alpha} \epsilon^{*\nu\sigma}\big(g_{\mu\nu}((p_\gamma p_\psi)\epsilon^{\psi*}_\alpha -(p_\gamma \epsilon^{*}_\psi)p^\psi_\alpha)((p_\gamma
p_\psi)\epsilon^{\psi}_\sigma-(p_\gamma \epsilon_\psi)p^\psi_\sigma) \\ \label{a1}
&+p^\gamma_\mu p^\gamma_\nu(M_\psi^2 \epsilon^{*\psi}_\alpha \epsilon^{\psi}_\sigma-p^\psi_\alpha p^\psi_\sigma)\big) \; ,
\end{align}
where the normalisation is given by
\begin{equation}
|A^{{\rm norm}}_2|^2 \sim \epsilon^{\mu}_{\phantom{\mu}\alpha} \epsilon^{*\alpha\nu}((p_\gamma p_\psi)^2g_{\mu\nu}+M_\chi^2 p^\gamma_\mu p^\gamma_\nu)+2(p_\gamma p_\psi)p^\gamma_\mu p^\gamma_\nu \,{\rm Re}[\epsilon^{\mu}_{\phantom{\mu}\alpha} \epsilon^{*\alpha\nu}] \; .
\end{equation}

G.~Anelli {\it et al.}  [TOTEM Collaboration],
    JINST {\bf 3} (2008) S08007.

\bibitem{FSC}M.~Albrow {\it et al.}  [USCMS Collaboration],
   arXiv:0811.0120 [hep-ex];\\
J.~W.~Lamsa and R.~Orava,
   arXiv:0907.3847 [physics.acc-ph].

\bibitem{cc} G.~V.~Pakhlova,
  arXiv:0810.4114 [hep-ex];\\
  Phys.\ Atom.\ Nucl.\  {\bf 72} (2009) 482
  [Yad.\ Fiz.\  {\bf 72} (2009) 518].
  
\end{document}